\documentclass[10pt,journal]{IEEEtran}
\usepackage[caption=false,font=normalsize,labelfont=sf,textfont=sf]{subfig}
\usepackage{setspace}
\usepackage{supertabular,caption}

\usepackage{xspace}
\usepackage{algorithmic}
\usepackage{algorithm}
\usepackage{microtype}
\usepackage{color}
\usepackage{url}
\usepackage{lipsum}
\usepackage{fancyhdr}


\usepackage[acronym]{glossaries}
\usepackage{soul} 

\usepackage{array}

\newcolumntype{P}[1]{>{\centering\hspace{0pt}}p{#1}}
\newcolumntype{M}[1]{>{\centering\hspace{0pt}}m{#1}}
\newcolumntype{L}{>{\centering\arraybackslash}m{3cm}}
\usepackage{epstopdf}	

\usepackage{supertabular,booktabs}
\usepackage{bbm}
\usepackage{multirow}
\usepackage[usenames,dvipsnames]{xcolor}
\renewcommand{\arraystretch}{1.5}
\usepackage{etoolbox}
\usepackage{pbox}
\usepackage{tabu}	
\usepackage{filecontents}
\usepackage{enumerate}
\usepackage{textcomp}

\usepackage{notoccite}

\usepackage{colortbl}
\usepackage{graphicx}
\usepackage{fancyhdr}
\usepackage{notoccite}

\usepackage{colortbl}
\usepackage{graphicx}

\fancypagestyle{firststyle}{
	\fancyhf{}
	\fancyhead[L]{O. Kanhere, H.Poddar, and T. S. Rappaport, ``Calibration of NYURay for Ray Tracing using 28, 73, and 142 GHz Channel Measurements conducted in Indoor, Outdoor, and Factory Scenarios,'' \textit{accepted and to appear in IEEE Transactions on Antennas and Propagation}, pp. 1--16, 2024}     %

}

\begin{document}
	
	\title{
		Calibration of NYURay for Ray Tracing using 28, 73, and 142 GHz Channel Measurements conducted in Indoor, Outdoor, and Factory Scenarios}
	
	\author{\IEEEauthorblockN{Ojas Kanhere, Hitesh Poddar, and Theodore S. Rappaport\thanks{This work is supported by the NYU WIRELESS Industrial Affiliates Program and National Science Foundation (NSF) Grants: 1909206 and 2037845}\\}
	}

	\maketitle
	\thispagestyle{firststyle}

\begin{abstract}
Site-specific wireless channel simulations via ray tracers can be used to effectively study wireless, decreasing the need for extensive site-specific radio propagation measurements. To ensure that ray tracer simulations faithfully reproduce wireless channels, calibration of simulation results against real-world measurements is required. In this study we introduce NYURay, a 3D ray tracer specifically tailored for mmWave and sub-THz frequencies. To reliably generate site-specific wireless channel parameters, NYURay is calibrated using radio propagation measurements conducted at 28, 73, and 142 GHz in diverse scenarios such as outdoor areas, indoor offices, and factories. 

Traditional ray tracing calibration assumes angle-dependent reflection, requiring slow iterative optimization techniques with no closed form solution. We propose a simpler and quicker novel calibration method that assumes angle-independent reflection. The effectiveness of the proposed calibration approach is demonstrated using NYURay. When comparing the directional multipath power predicted by NYURay to the actual measured power, the standard deviation in error was less than 3 dB in indoor office environments and less than 2 dB in outdoor and factory environments. The root mean square (RMS) delay spread and angular spread was underpredicted by NYURay due to incomplete environmental maps available for calibration, however an overall agreement between the measured and simulated values was observed. These results highlight the high level of accuracy NYURay provides in generating the site-specific real-world wireless channel, that could be used to generate synthetic data for machine learning.
\end{abstract}


\IEEEpeerreviewmaketitle
\section{Introduction}
A wide range of new applications, including centimeter-level position location, high-resolution virtual/augmented reality (VR/AR), industrial automation, environment monitoring, and imaging, will be made possible by the enormous spectrum at mmWave and sub-THz frequencies \cite{Rappaport19a}. Significant efforts have been made to characterize the wireless channel in order to assess the viability of such applications. Moreover, for the purpose of creating accurate channel models, several channel measurements have been conducted at mmWave and sub-THz frequencies \cite{Ju_2021,Mac15b} in different environments at different frequency bands. However, real-world channel measurements are extremely time consuming, with typical measurement campaigns only gathering a handful of locations per environment. Furthermore, because of the sophisticated hardware needed, performing wireless channel measurements is frequently cost prohibitive. Thus, if there is an assurance that signal characteristics of predicted wireless channels, such as the powers, angle of arrival (AoA), angle of departure (AoD), and time of flight (ToF) of signals arriving at wireless receivers closely match the true measured signal characteristics at given locations, site-specific simulation-based prediction of wireless channels can supplement actual measurements.
\par Based on the position of the wireless transmitter (TX), receiver (RX), and environmental obstacles ray tracing is a potent tool that may be utilized in predicting the signal propagation characteristics of wireless signals. When a signal is propagating, obstructions in the environment could cause reflections, transmissions, diffraction, or scattering. The ray tracer makes predictions about the wireless signals' strength, AoA, AoD, and ToF, which can be used to recreate the spatial and temporal properties of the wireless channel for simulation-based study. Furthermore, ray tracing can be used to generate synthetic data to train machine learning algorithms and to evaluate sensing applications such as position localization and seeing through walls. At locations in a particular environment where the channel measurements were not made, artificial data can be produced to assess how well the sensing algorithms are performing. Ray tracers can offer precise ToF data that sensing algorithms need but may not have gathered during wireless channel measurements \cite{Kanhere19a}.
\par Although ray tracing as discussed in this paper focuses on applications for wireless communications, ray tracing originated in computer graphics, wherein the lighting of a scene is predicted, i.e., the reflections, refraction, diffraction, and scattering of visible light originating from a light source in an environment is predicted.
In contrast, ray tracing in the context of wireless communications has been used for optimal TX placement for indoor wireless systems \cite{Verstak_2002} to maximize signal coverage in the environment and minimize the bit error rate (BER). Additionally, ray tracing has been used to simulate the performance of applications such as position location \cite{Kanhere19a}. 
\par However, using a ray tracer for wireless simulations ``out of the box" could result in inaccurate predictions of the wireless channel because the predicted arriving signal levels for a specific TX-RX pair may differ from the signals that would have been measured with the channel sounder at the same TX-RX location. For instance, an overestimation of the channel capacity could result from a ray tracer that predicts increased spatial diversity in the wireless channel. If employed for network planning, inaccurate power predictions made by the ray tracer could result in suboptimal base station placement that would cause signal outages in certain areas. For ray tracer-assisted simulations, accurate reproduction of the site-specific wireless channel requires proper calibration of the ray tracer.
\par This paper is organized as follows. Section I presents an overview of the calibration techniques used for calibrating ray tracer for wireless communication. The popularly used techniques for wireless ray tracing, their computational complexity and the hybrid ray tracing technique adopted in NYURay is introduced in section II. In Section III, we present the theoretical propagation models used by NYURay, a 3-D mmWave and sub-THz ray tracer developed at NYU WIRELESS, to simulate reflection, penetration and scattering of wireless signals for site-specific wireless channel prediction \cite{Kanhere_2022}. Additionally, in Section IV a detailed insight is provided about the measurement campaigns and the channel sounder used in these measurement campaigns to conduct channel measurements at 28, 73 and 142 GHz in outdoor, indoor and factory scenarios. Furthermore, Section V describes how NYURay is used to analyze the channel measurements. Moreover, in Section VI a description of the procedure that can be used to calibrate NYURay to directional channel measurements conducted in a variety of indoor and outdoor environments at 28, 73, and 142 GHz is provided (over a total of nine different measurement campaigns). 
In Section VII, we present a comprehensive analysis that compares the power, RMS angular spread, and RMS delay spread at 28, 73, and 142 GHz for outdoor, indoor, and factory scenarios using the channel measurements and the predictions obtained through NYURay \cite{Kanhere_2022}. Additionally, in Section VIII the electrical properties of building materials, as obtained from the calibration technique used in NYURay, are provided \cite{Kanhere_2022}. Finally, conclusions are provided in Section IX.

\begin{table}[h!]
	\centering
		\caption{Table of Mathematical Notations}
	\begin{tabular}{| m{1.5cm} | m{6cm} |}
		\hline
		\textbf{Symbol} & \textbf{Description} \\
		\hline
        \( \binom{n}{r} \) & Binomial coefficient \\
		\hline
$ \mathcal{O} $ &  Big O notation, describes an upper bound on the time complexity of an algorithm\\
		\hline
		x & Scalar value x\\
		\hline
		\textbf{x} & Vector x\\
			\hline
	$ |\textbf{x}| $& Magnitude of x \\
	\hline
$ 	\textbf{x}^T $& Transpose of vector x\\
\hline
$A^{-1}$& Inverse of matrix A\\
		\hline
				$\sum$ & Summation, the addition of a sequence of numbers\\
		\hline
		$\prod$ & Product, the multiplication of a sequence of numbers \\
		\hline
	\end{tabular}
	\label{tab:math_notations}
\end{table}

\section{Prior mmWave and sub-THz Ray Tracers and Metrics for Calibration}
mmWave ray tracers have been created by a variety of research organizations and companies for use in environments that are indoors as well as outdoor. Channel statistics like path loss, RMS angular spread, and RMS delay spread are frequently used for comparison in order to assess the performance of mmWave ray tracers. Within an Urban Microcell (UMi) environment at 28 GHz, authors in \cite{Lee_2018} examined the measured and estimated omnidirectional path loss across various LOS and NLOS conditions. Moreover, in order to reduce the error in route loss prediction in an indoor office environment \cite{Liu_2017} and an outdoor UMi environment \cite{Wei_2017} at 28 GHz, researchers fine-tuned a single parameter of the relative permittivity of materials in the environment. In \cite{Hur_2016}, the authors linearly interpolated the dielectric constant of concrete at 5.2 GHz and 60 GHz, to obtain the electrical characteristics of concrete at 28 GHz. The authors assumed all buildings were made of concrete \cite{Hur_2016}. An empirical ray tracing model was used in simulations at 73 GHz in \cite{Nguyen_2014}, wherein the reflection loss varied linearly with incident angle. The measured and simulated RMS angular and RMS delay spreads were compared in urban street canyon and UMi environments at 28 GHz and 73 GHz in \cite{Hur_2016} and \cite{Nguyen_2014} respectively. The path loss exponent (PLE), obtained via least squares estimation across multiple TX-RX locations was also compared in \cite{Hur_2016} and \cite{Nguyen_2014}, however the difference in path loss at individual TX-RX locations was not compared. Additionally, in \cite{Karstensen_2016}, the power-angular-delay profiles of a huge unoccupied room and a tiny workplace at 29 GHz (for a single TX-RX site pair in both situations) were compared. The ray tracer's ability to forecast the dominating specular pathways was demonstrated, but a quantitative analysis of the differences between the measured channel and the simulated channel was omitted. At 60 GHz in an interior office space, the route gains of various multipath components were contrasted in \cite{Zhou_2017_RT}. The direct comparison of simulated directional power to measured directional power, for example, is a more precise comparison that is rarely made.
\par Several ray tracers operating at sub-THz frequencies have also been developed. Often the material parameters up to 100 GHz are extrapolated to sub-THz frequencies \cite{Gougeon_2019}. In \cite{Piesiewicz_2007,Han_2015}, the authors measured the refractive index of material such as wallpaper and plaster via THz time-domain spectroscopy and calculated the reflection coefficients via Fresnel's equations, observing good agreement. However, the reflection and scattering parameters were directly used for ray tracer, without validation in more complex environments. Kurner et al. \cite{Kurner_2012,Priebe_2013} calibrated an indoor sub-THz ray tracer at 300 GHz to directional measurements conducted at TX-RX locations by minimizing the difference in power of the measured and simulated rays. Work in \cite{Guan_2019} characterized the train-to-infrastructure (T2I) channel in a train station, at 300 GHz and provided ray tracing results at two TX-RX locations with TX-RX separations of 5.2 m and 6.7 m. In \cite{Guan_2019b}, the authors characterized the wireless channel inside a train wagon at 60 and 300 GHz, and calibrated an in-house ray tracer to the path gain, delay, and AoA/AoD of significant measurement rays. Ray tracing simulations in a corridor were validated for LOS locations in \cite{Sheikh_2022} at 90, 95, and 100 GHz by comparing the measured and simulated RX power. In \cite{Dupleich_2020}, the authors presented measurements conducted at 190 GHz in a conference room and compared the overall path loss, angular and delay statistics of six measured TX-RX links versus more than 15000 simulated positions in the same environment.

\section{Ray Tracing Computational Methods}
Ray tracing computations for specular reflections and penetration are primarily conducted via two methods - image-based ray tracing (ray tracing by the method of images) and using the shooting-bouncing rays (SBR) technique.

Image-based ray tracing \cite{McKown91a,Ho94a} relies on the principle that for a ray impinging an obstruction along the path of the ray from the TX to the RX, the incident ray and the reflected ray make the same angle with the normal to the plane containing the obstruction. Obstructions are treated as infinitely long, thin mirrors when using image-based ray tracing, which is justified for regular-sized obstructions in the environment at mmWave and sub-THz frequencies, since the short ($<$1 mm) wavelengths at these frequencies make the obstructions electrically large \cite{Ho94a}. The method of images is used to calculate the precise signal path from the TX to the RX. In the method of images, a virtual RX is placed at the mirror image of the true RX position, to model reflections from the obstructions in the environment. The images of the RX are calculated successively across at most $ k $ obstructions, where $ k $ is the maximum number of reflections a ray may go through~\cite{Ho94a}.

The shortcoming of image-based ray tracing is that if there are a large number of obstructions, the simulation run-time is large. Although the image-based ray tracing method finds the direction of arrival of rays very accurately, finding the reflection of the RX, recursively, from all combinations of obstructions is computationally expensive. Assuming that each ray is reflected at most $ n $ times, with $ N_o $ obstructions in the environment, there are ${N_o\choose n}$ images that need to be computed ($ \mathcal{O}(N_o^n) $) \cite{Kanhere19a}. Thus, the computational complexity grows exponentially with the number of obstructions in the environment \cite{Durgin97a}.

On the other hand, a ray tracer using the SBR technique, launches rays uniformly in all directions and then traces the path of each launched ray, as the ray interacts with various obstructions in the environment. Each launched ray represents a continuous wavefront and carries the power that would have been carried by the wavefront \cite{Durgin97a}. 

For launched rays to represent equal portions of the wavefront, the rays must be uniformly distributed around a unit sphere, centered at the TX. One method to ensure the uniform sampling of the unit sphere is to use a tessellated icosahedron centered at the TX. An icosahedron is a twenty-faced regular polygon with triangular faces. To tessellate an icosahedron, each triangular face is subdivided into equilateral triangles. 
The tessellation frequency $ N_t $ is the number of such equilateral triangles constructed on each edge of the icosahedron. As derived in \cite{Durgin97a}, rays launched from the center of the icosahedron to the vertices of the so-formed equilateral triangles have approximately equal angular spacing ($ \alpha $) given by:
\begin{align}
	\alpha = \dfrac{69^\circ}{N_t}.
\end{align}
In total, with a tessellation frequency of $ N_t $, the tessellated icosahedron has $N_v =  10N_t^2+2$ vertices. 

To determine the rays that reach the RX, the reception sphere is used \cite{Durgin97a,Honcharenko_1992,Schaubach92a}. The radius of the reception sphere ($ r $) is proportional to the $ToF$ of the ray and is given by:
\begin{align}
	r = \dfrac{\alpha d}{\sqrt{3}} \label{eq:d_sphere_r}
\end{align}
where $ d = c \times ToF $ is the total propagation path length of the ray. Rays that arrive within the reception sphere are said to have reached the RX. Setting the radius of the reception sphere as in \eqref{eq:d_sphere_r} ensures that a single ray from each wave front is captured by the reception sphere. 

The computational complexity of SBR ray tracing is lower than the complexity of image-based ray tracing. For simulating the first-order (single-bounce) interactions (reflections and transmissions), $ N_vN_o $ ray-object intersections were computed (one intersection test between each of the $N_v$ rays and each of the $N_o$ obstructions was computed ). Further, $ N_v $ reflections must be computed. Single-bounce object interactions of $ N_v $ rays can generate at most $ 2N_v $ rays (each ray may be reflected and/or transmitted through the first obstruction in the path of the ray). Finally, ray reception tests of the $ 2 N_v $ rays (using the reception sphere) must be carried out. Extending the above analysis to multi-bounce interactions, assuming a maximum of n ray-object interactions, the number of ray-object intersection tests are $ N_vNo + 2 N_vNo + \cdots  2^{n-1}N_vNo = N_vNo( 2^n-1)$, the number of reflections to be computed are $N_v +2N_v +4N_v \dotsc + 2^{n-1}N_v=N_v(2^n-1)$, and the number of ray reception tests to be done are $  2N_v + 4N_v +\dotsc 2^{n}N_v=2N_v(2^n-1)$. The above analysis provides an upper bound to the computation required for SBR, since typically a majority of rays are dropped at each interaction order since the predicted power falls below the RX sensitivity. In general, the computational complexity of SBR scales linearly with the number of obstructions in the environment ($ \mathcal{O}(N_o) $) \cite{Durgin97a}. 

The accuracy of the AoA of rays received at the RX depends on the number of rays launched from the TX. For example, if rays are launched from the vertices of a tessellated icosahedron with $ N_{t}$, since the average radial separation between two rays is $ \dfrac{69^\circ}{N_{t}} $ \cite{Durgin97a}, for sub-degree accuracy for AoA, $ N_{t}>50 $, which is computationally expensive. By increasing $N_{t}$, a greater number of rays may be launched, however the computational complexity of the ray tracer increases as more rays are launched leading to longer simulation times.

To reduce computational overhead, NYURay uses a hybrid ray tracing algorithm \cite{Tan96a}, which combines the shooting-bouncing rays (SBR) ray tracing technique \cite{Schaubach92a, Durgin97a} with image-based ray tracing \cite{McKown91a,Ho94a}. Since rays are launched at discrete angles (with an angular separation of $ \alpha $) via SBR ray tracing, the trajectories of rays that reach the RX are approximate trajectories, with an angular error of up to $ \alpha $. Nevertheless, when the icosahedral tessellation frequency ($ N_t $) is sufficiently large, despite potential errors in the computations of the points of ray incidence, the sequence of reflecting surfaces is reliably determined. Recursive images of the RX are then computed at the reflecting surfaces determined. Finally, the ray trajectory is accurately calculated by connecting all the RX images. The computational complexity of the hybrid ray tracing algorithm scales linearly with the number of obstructions in the environment ($ \mathcal{O}(N_o) $) \cite{Durgin97a}, since beyond the standard SBR ray tracing computations, the additional computation required in calculating path corrections for $ N $ received rays ($ N<<N_o $) is minimal.

In contrast to reflections and penetration (which may be computed via image-based ray tracing or SBR) and scattering (which may be computed via SBR), diffraction is often computed using either the Knife Edge Diffraction (KED) model or the Uniform Theory of Diffraction (UTD) \cite{Tan96a}. The KED model approximates diffraction effects by modeling obstructions as a single vertical strip. The UTD model accounts for multiple diffracting edges and complex geometries.

\section{Ray Propagation Mechanisms}

Changes in the direction of propagation and the energy carried by the signal occur as a result of objects in the environment interacting with signals as the signal propagates through the wireless medium. Reflection and transmission are the predominant modes of propagation at mmWave and sub-THz frequencies. Diffused scattering also contributes significantly to propagation in areas where specular reflections from the TX are inhibited. However, in the presence of specular reflections, the power carried by scattering waves at mmWave and sub-THz frequencies is negligible (as much as 30 dB lower than the power contained in specular reflections \cite{Jansen_2011,Ju19a,Taleb_2023}). In \cite{Priebe_2011}, authors show that the difference in channel transfer functions with and without scattering is less than a dB. 

Diffraction is not a dominant propagation mechanism is negligible at mmWave and sub-THz frequencies. In \cite{Rappaport_2017}, based on indoor and outdoor measurements at mmWave frequencies, the authors concluded that mmWave diffraction will not be a dominant propagation mechanism. In \cite{Jacob_2012,Kurner_2012} the authors show that the median difference in path loss by not considering diffraction was less than 0.05 dB, and the 90 percentile difference in path loss was less than a dB for a wooden wedge and a metallic wedge at 300 GHz. The authors in \cite{Jacob_2012} concluded that diffraction can be neglected almost everywhere in a room. The authors in \cite{Moraitis_2023} show that for very shallow angles ( 5\textdegree ), a diffraction gain of up to 2.1 dB is possible, however in general, reflection is the dominating propagation mechanism and diffraction can be neglected.

Therefore, neglecting diffraction, NYURay models propagation due to specular reflections, transmissions, and diffuse scattering. Further, during calibration of specular reflection and transmission properties of materials, we assume that the scattered multipath components are much lower than specular multipath components.

\par
\begin{figure}[]
	\centering
	\includegraphics[width=0.4\textwidth]{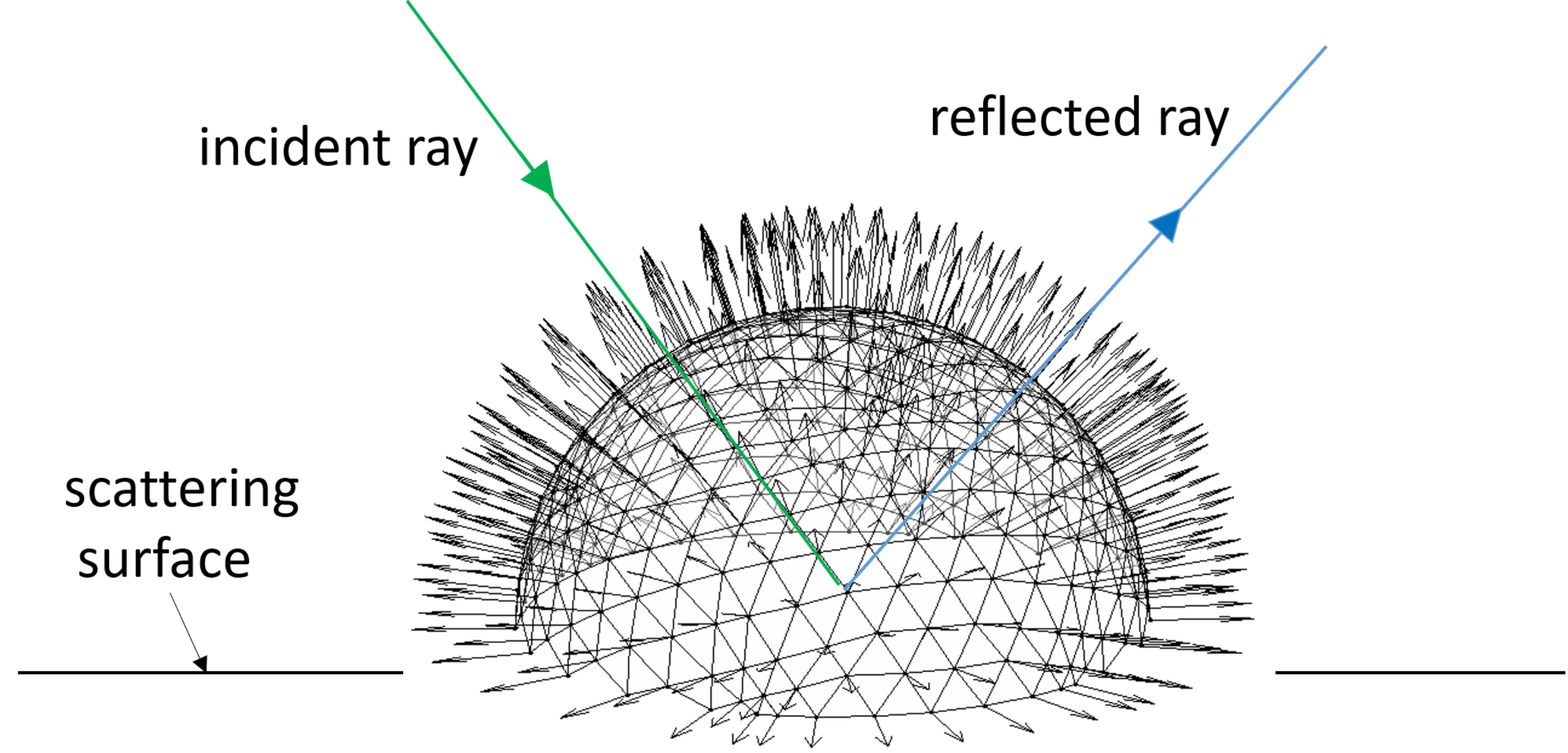}
	\caption{When a ray is incident on a scattering surface, NYURay generates scattering rays (represented by black arrows) that propagate along the vertices of a half-icosahedron.}
	\label{fig:half_ico}
\end{figure}

 In every direction where a ray was launched, on encountering an intersection with an obstruction, three types of rays are created for further propagation analysis - the specular reflected ray, the scattered rays, and the transmitted ray. By Snell's Law, the reflected ray and the incident ray make equal angles with the normal to the obstruction, while the transmitted ray propagates in the same direction as the incident ray, with a lateral displacement. However, signals at mmWave and sub-THz frequencies do not penetrate through walls with significant thickness, such as brick and concrete walls. Hence a ``thin wall” assumption is made in this article, wherein the signals are assumed to propagate along the same path as the incident ray. Scattered rays are launched from the vertices of a half-icosahedron, centered at the point of incidence of the signal on the obstruction, as seen in Fig.\ref{fig:half_ico} . The strongest scattered signals are simulated close to the direction of the specular reflection. The number of scattered rays, equal to the number of vertices of the half-icosahedron, is a user-controlled number.
\subsection{Signal Reflection}
The reflected signal is assumed to follow Snell's law at the point of incidence, i.e., the angle of incidence of the ray at the obstruction is equal to the angle of reflection. New source rays at each boundary are then recursively traced in the reflection direction to the next encountered obstruction on the propagating ray path. 

In the literature, reflection loss is often modeled using Fresnel's equations\cite{Rap02a}. However, due to the highly non-linear nature of the equations, a closed form solution to minimizing the difference in the measured multipath component powers to the simulated ray tracer output to field measurements is not attainable. To reduce the complexity of the problem, researchers often choose to either assume that all objects in the environment possess the same relative permittivity \cite{Liu_2017,Wei_2017,Hur_2016}, or use extrapolated values of permittivity that were obtained from sub-6 GHz measurements that may not be accurate at higher frequencies \cite{Gougeon_2019}.

Taking inspiration from prior work in \cite{Durgin_1998}, we propose the use of a simplified and more analytically tractable reflection loss model, wherein the reflection loss of a ray impinging an obstruction is assumed to be constant, independent of incident angle. As shall be explained in Section \ref{sec:calibration}, such simplification allows for a closed-form solution for minimizing the difference in the measured and simulated path gains, allowing quick and accurate calibration of the electrical properties of the materials in the environment. Work in \cite{Durgin_1998} derived a least-squares formulation to determine partition losses of objects lying between the TX and RX. In this work, we shall derive a least-square formulation to determine both the reflection coefficients and the partition losses of obstructions in the signal path. Thus, the reflection propagation mechanism is simplified to an angle independent constant reflection loss model as follows:

\begin{align}
	\dfrac{|E_r|}{|E_i|} =|\Gamma_m| ,
\end{align}
where $ |E_r| $ is the magnitude of the reflected electric field, $| E_i| $ is the magnitude of the incident electric field, and $|\Gamma_m|$ is the magnitude of the reflection coefficient of the obstruction. Thus, the reflected power $ P_r = |\Gamma_m|^2 P_i$, where $ P_i $ is the power incident on the obstruction. Written in logarithmic units (dB), the reflected power in terms of incident power is given by:
\begin{align}
	P_r [dBm] = P_i[dBm] - L_{r,m}[dB]
\end{align}  
where $ L_{r,m} $ is the reflection loss of the obstruction $m$, and is equal to $ 20 \log_{10}(|\Gamma_m|) $.

Although Fresnel equations are derived from fundamental electromagnetic principles, and hence have a physical basis in the prediction of reflected power, in the absence of knowledge of the complex environment where measurements were conducted, the measured reflection loss may not be modeled via Fresnel's equations. As seen in \cite{Choi_2015}, no clear dependence between angle of incidence and reflected power may be observed, if insufficient information of the material being tested (such as fine-grain details of the material under test) is available. In situations where a detailed environmental map is unavailable for ray tracer calibration, a simplified reflection loss model can be used to model reflections and accurately predict the power contained in arriving multipath components \cite{Schaubach92a}. 

Reflections of order up to the fifth order (i.e., up to fifth bounce reflections along a propagation path) were considered for ray tracing simulations, since higher order reflections were not observed in the field measurements at 28, 73, and 142 GHz \cite{Rap13a}.

\subsection{Signal Penetration}
At mmWave and sub-THz frequencies, although signal penetration through materials like external building walls composed of granite, brick, and metal is infeasible, certain objects such as wood, drywall, and glass allow signal to propagate through. 

The wall attenuation factor (WAF) model \cite{Seidel_1992,Durgin_1998} has been used to model the signal partition loss through materials as follows:
\begin{align}
	P_t [dBm] = P_i [dBm] - L_{t,m} [dB]
\end{align}

where $ P_t $ is the power of the transmitted signal in dBm, $ P_i $ is the power of the incident signal in dBm, and $ L_{t,m} $ is the partition loss of obstruction $ m $.
\subsection{Scattering}
For a smooth surface, the incident wave reflects off the surface in the specular direction as determined by Snell's law. For rough objects, the incident signal gets re-radiated in all directions due to \textit{scattering} \cite{Rap02a}.

The roughness of an object is tested by the Rayleigh criteria, which defines a critical height of surface perturbations, $ h_c $, as follows:
\begin{align}
	h_c = \dfrac{\lambda}{8 \cos(\theta_i)}
\end{align}
where $ \lambda $ is the wavelength of the radio frequency (RF) signal and $ \theta_i  $ is the angle of incidence. If the range of surface perturbations is less than $ h_c $, the surface is considered smooth. At sub-6 GHz frequencies, objects are relatively smooth (with a surface roughness smaller than the wavelength of the incident wave) due to which scattering is a much weaker propagation mechanism than reflection and is typically ignored in wireless communication design and simulations. In contrast, at mmWave and sub-THz frequencies, since the carrier wavelength is in the order of the surface perturbations of surfaces, the power carried by scattered rays needs to be considered, especially in NLOS conditions where the scattered rays may act as the dominant propagation mode in the absence of specular reflections \cite{Rap02a}.

To model the power carrier by the scattering rays, the directive scattering (DS) dual lobe model \cite{Degli_Eposti2007} was employed, which was shown to match scattering patterns observed at mmWave and sub-THz frequencies in \cite{Ju19a}. The dual lobe model includes both forward scattering and backscattering.
According to the dual-lobe DS model the scattered electric field $E_s$ is given by:

\begin{eqnarray}~\label{equ:dual}
	|E_s|^2=|E_{s0}|^2 \Bigg[\Lambda\cdot \left(\dfrac{1+\cos(\Psi_i)}{2} \right)^{\alpha_R}\nonumber\\
	+ (1-\Lambda)\cdot \left(\dfrac{1+\cos(\Psi_i)}{2} \right)^{\alpha_i}\Bigg] 
\end{eqnarray}
where $E_{s0}$ is the maximum scattered electric field, $\Psi_i$ is the angle between the scatter ray and the incident ray, $\Lambda$ determines the relative strength of the back-scatter lobe with respect to the forward-scatter lobe, and $\alpha_i$ and $\alpha_R$ are the back-scattering and forward-scattering lobe width parameters, respectively \cite{Degli_Eposti2007}.

\subsection{Propagation Loss}
In addition to the reflection and partition loss in power experienced by signals at obstructions, a free space propagation loss with a $ 1/d^2 $ dependence of received power on the propagation distance was assumed, calculated based on Friis' free space path loss (FSPL) model \cite{Sun16a}, where $ d $ is equal to the total propagated ray length. 

For diffused scattering with path segments having lengths $ s_1 $ and $ s_2 $, a power dependence proportional to $ 1/(s_1s_2)^2 $ was assumed, to account for signal spreading of the scattered signal \cite{Seidel94a}.

\section{mmWave and sub-THz Channel Measurements}\label{sec:measurements}

NYU WIRELESS has performed mmWave and sub-THz channel measurements at 28, 73, and 142 GHz over a ten-year period (2012-2022), in and around the Brooklyn and Manhattan campuses of NYU \cite{Rap13a,Ghosh_2014,Sulyman_2014,Rappaport_2015,Maccartney_2015,Xing21a,Xing_2021d}, in the indoor office environment of the 9$ ^{th} $ floor of 2 Metrotech (the floor where the laboratories of NYU WIRELESS were located prior to 2019) \cite{Mac15b,Ju_2021}, and in four different factory environments across New York (NY) City \cite{Ju_2022,Ju_2024a}. The measurements made by graduate students at NYU WIRELESS resulted in the creation of channel models for mmWave and sub-THz communications, demonstrating the viability of these frequencies \cite{Rap13a,Rappaport19a}. The environments where the mmWave and sub-THz channel measurements were conducted are described in this section. Additionally, the hardware that was used to carry out the channel measurements is also introduced. Three generations of channel sounders were developed at NYU WIRELESS, starting with the first mmWave time-domain based sliding-correlation channel sounder operating at 28 GHz, followed by the mmWave channel sounder at 73 GHz, and the latest sub-THz channel sounder at 142 GHz.

\subsection{Measurement equipment}\label{sec:equipment}
The channel measurements were conducted using sliding-correlation based channel sounders operating at 28, 73, and 142 GHz. All three channel sounders had similar super-heterodyne RF architectures, wherein a pseudo-random noise (PN) sequence of 2047 chips in length (11 bits) was modulated to an intermediate frequency (IF) and then upconverted to the desired RF after mixing with a local oscillator in an RF upconverter. 

\begin{figure}[]
	\centering
	\includegraphics[width=0.35\textwidth]{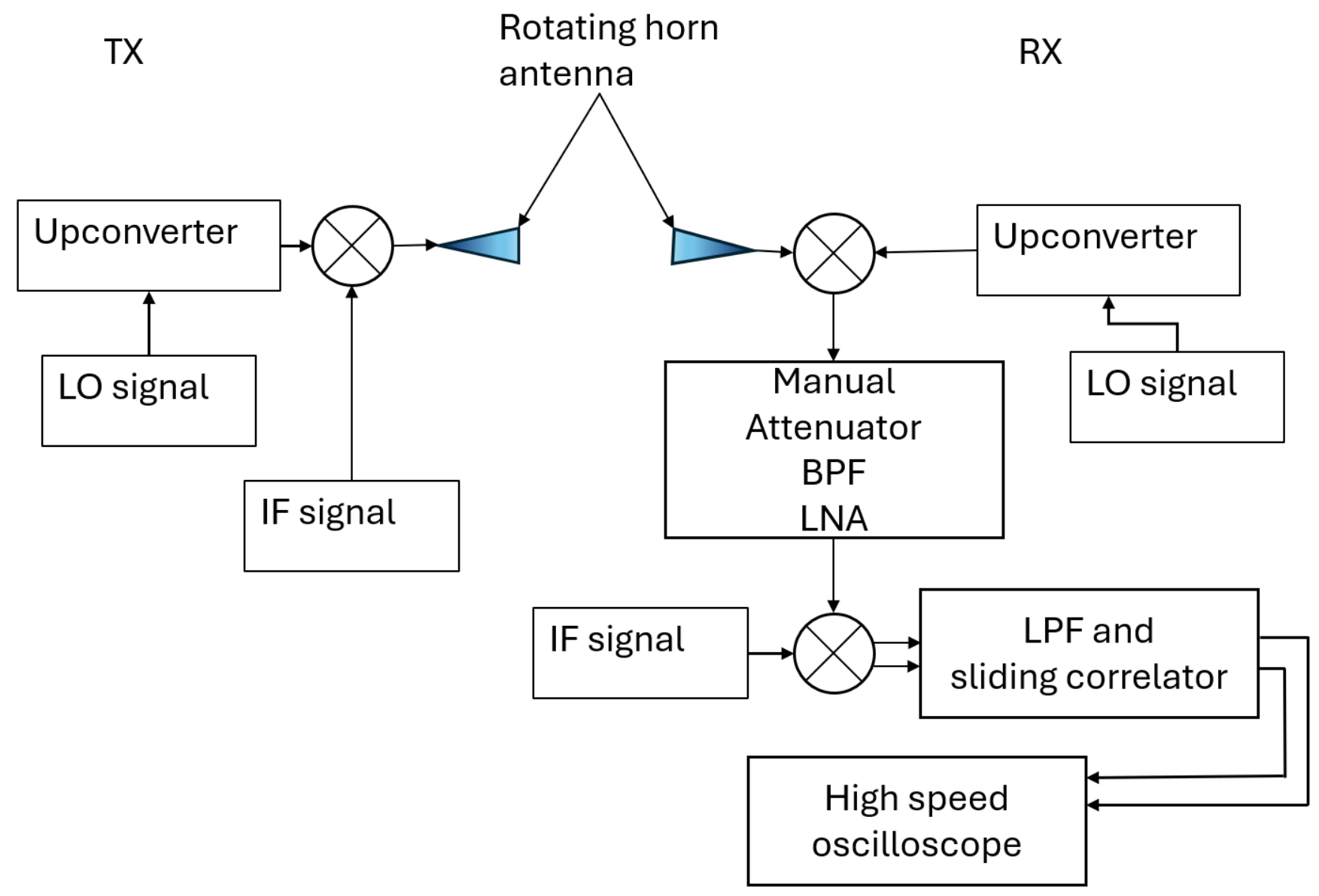}
	\caption{Block diagram of channel sounder used for propagation measurements. Details of the LO signals, IF signals, filters and upconverters used are provided in \cite{Rap13a,Mac17a,Xing19a} }
	\label{fig:block_diagram}
\end{figure}

The RF signal was transmitted out of a high-gain directional horn antenna mounted on an electronically controllable gimbal, capable of rotating in the azimuth and elevation plane. The RF signal is analogously received by high-gain directional horn antenna mounted on an identical gimbal at the RX. The RF signal at the RX is downconverted by an local oscillator (LO) signal, passed through a low noise amplifier (LNA) and is IQ demodulated to baseband. Details of the environment, RF bandwidths, TX and RX horn antenna half-power beamwidths, and gains of the TX and RX horn antennas used at each of the three frequencies (28, 73, and 142 GHz) are summarized in Table \ref{tbl:hardware}. A block diagram of the channel sounder is provided in Fig. \ref{fig:block_diagram}

	\renewcommand{\arraystretch}{1.2}
	\begin{table*}[]\caption{\centering The environment, RF bandwidths, TX antenna HPBW beamwidth and gain, and RX antenna HPBW beamwidth and gain used in channel measurements at 28, 73, and 142 GHz \protect{\cite{Ju_2021,Mac15b,Rap13a,Rappaport_2015,Mac14a,MacCartney_2019,Xing_2018,Xing21a,Ju_2022b, Shakya_2023}}. }\label{tbl:hardware}
		\centering
	
		\small
		\begin{tabular}{|c|c|c|c|c|c|c|}
			\hline
			\begin{tabular}[c]{@{}c@{}} \textbf{Frequency}\end{tabular}           & \textbf{Environment}   & \begin{tabular}[c]{@{}c@{}}\textbf{RF}\\ \textbf{Bandwidth}\end{tabular} &\begin{tabular}[c]{@{}c@{}}\textbf{TX Antenna}\\ \textbf{HPBW}\end{tabular} & \begin{tabular}[c]{@{}c@{}}\textbf{TX Antenna}\\ \textbf{Gain}\end{tabular} & \begin{tabular}[c]{@{}c@{}}\textbf{RX Antenna}\\ \textbf{HPBW}\end{tabular} & \begin{tabular}[c]{@{}c@{}}\textbf{RX Antenna} \\ \textbf{Gain}\end{tabular} \\\hline
			\multirow{2}{*}{28 GHz}  & Indoor Office &     800 MHz         &      30\textdegree                                                     &    15 dBi                                                       &     30\textdegree                                                       &        15 dBi      \cite{Mac15b}                                               \\\cline{2-7}
			& Outdoor       &     800 MHz         &       10.9\textdegree                                                    &  24.5 dBi                                                         &       10.9\textdegree                                                    &       24.5 dBi                                                 \cite{Rap13a,Rappaport_2015}    \\\hline
			73 GHz                   & Outdoor       &  1 GHz            &     7\textdegree                                                      &         27 dBi                                                  &       15\textdegree                                                    &            20 dBi                             \cite{Mac14a,MacCartney_2019}                   \\\hline
			\multirow{3}{*}{142 GHz} & Indoor Office &   1 GHz           &   8\textdegree                                                        &   27 dBi                                                        &      8\textdegree                                                     &             27 dBi                                         \cite{Ju_2021,Xing_2018}      \\\cline{2-7}
			& Outdoor       &     1 GHz         &   8\textdegree                                                        &   27 dBi                                                        &      8\textdegree                                                     &             27 dBi \cite{Xing21a, Shakya_2023}\\\cline{2-7}
			& Factory       &      1 GHz        &      8\textdegree                                                        &   27 dBi                                                        &      8\textdegree                                                     &             27 dBi    \cite{Ju_2022b,Ju_2024a}   \\\hline                                                   
		\end{tabular}
		
	\end{table*}

To capture all arriving multipath propagation in the environment, the TX and RX gimbals were rotated in the azimuth and elevation plane. The gimbal may be swept 360\textdegree\ in the azimuth and 120\textdegree\ in the elevation plane in 1\textdegree\ steps, allowing double directional channel measurements (i.e., both AoA and AoD information was obtained from the channel measurements). A channel power-delay-profile (PDP) was captured at different TX-RX antenna pointing combinations. The channel sounder used for channel measurements at 28 GHz had an RF bandwidth of 800 MHz, while the channel sounders at 73 and 142 GHz utilized an RF bandwidth of 1 GHz. Multipath components separated by less than the PN sequence chip width could not be separately resolved, thus a multipath component resolution time of 2.5 ns was possible at 28 GHz while a slightly lower resolution of 2 ns was achieved at 73 and 142 GHz. The time resolution of the measurement system limits the multipath-resolving capability of the system, since multipath components arriving within the time resolution of the channel sounder are indistinguishable, if arriving from the same direction. Multipath components that arrived within the time resolution of the channel sounder, from different directions, were resolvable due to the highly directional horn antennas that acted as spatial filters, suppressing multipath components not arriving from the direction in which the TX and RX antennas were pointed in the directional measurements. If the difference in propagation distance of two multipath components were less than 75 cm in the 28 GHz measurement campaign (60 cm in the 142 GHz measurement campaign), the difference in arrival times of the multipath components would be 2.5 ns (2 ns in the 142 GHz measurement campaign). Thus, two multipath components with path lengths differing by less than 75 cm at 28 GHz and 60 cm at 142 GHz would be indistinguishable, if arriving from the same direction. 

The 2D horn antenna patterns were characterized in a free-space indoor laboratory environment by rotating the RX in the azimuth, with a TX-RX separation distance of 4 m (the RX was in far-field). The relative gains of the horn antennas were measured at various pointing angles, with the gain at boresight obtained from the manufacturer. In total, five types of horn antennas (based on five distinct TX and RX horn antenna HPBW) were used for the nine measurement campaigns, as is indicated in Table \ref{tbl:hardware}. 

\begin{figure}[]
	\centering
	\includegraphics[width=0.35\textwidth]{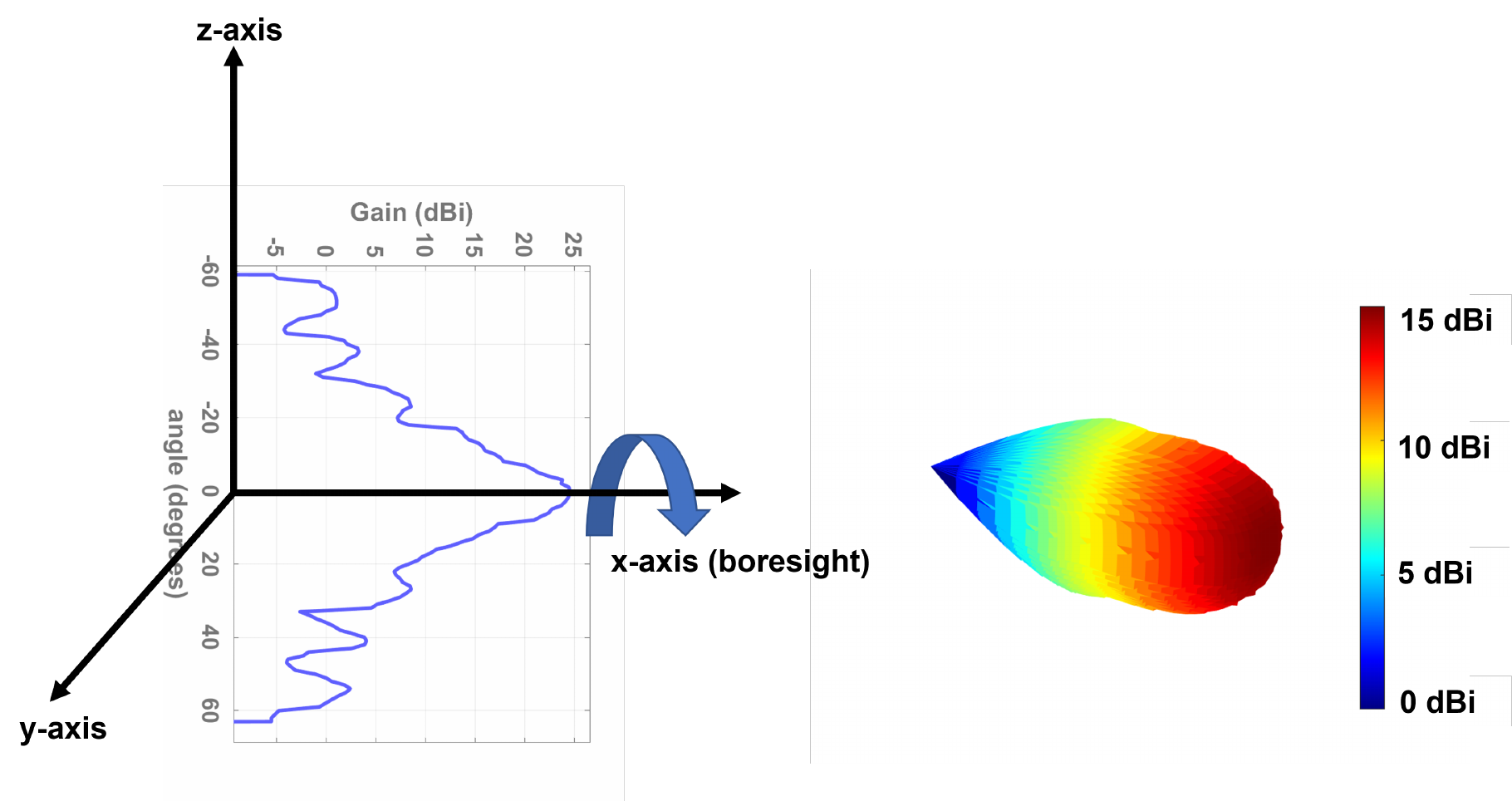}
	\caption{The 3-D antenna pattern of the horn antenna with a HPBW of 10.9\textdegree\ operating at 28 GHz is obtained by rotating the 2-D horn antenna pattern about the boresight axis (the x-axis). A symmetric antenna pattern was assumed in the vertical plane.}
	\label{fig:create_3D_antenna}
\end{figure}

Since the horn antennas had similar antenna pattern in the E and H planes, to transform the 2D antenna patterns to 3D antenna patterns, rotational symmetry about the boresight axis of the antenna was assumed. The antenna pattern, measured in the azimuth plane, was rotated about the boresight direction of the horn antenna to generate the 3D antenna pattern. As seen, for example, in Fig. \ref{fig:create_3D_antenna}, the 2-D antenna pattern of the horn antenna operating at 28 GHz is rotated about the x-axis, which aligns with the boresight direction of the horn antenna. 

\subsection{Measurement Campaigns and Environments}
The measurement equipment based on sliding-correlation channel sounding was used to characterize the mmWave and sub-THz wireless channel in three different environments - the indoor office environment (at 28 and 142 GHz), the outdoor urban microcell (UMi) environment (at 28, 73, and 142 GHz) and the factory environment (at 142 GHz), with the hardware used in each environment at each frequency listed in Table \ref{tbl:hardware}. In this paper, we define a measurement campaign to be a set of channel measurements collected at a given location, at a particular frequency. A total of nine measurement campaigns at mmWave and sub-THz frequencies were used to calibrate NYURay. Two measurement campaigns were conducted in an indoor office environment at 28 and 142 GHz \cite{Mac15b,Ju_2021}, three campaigns were conducted in the outdoor UMi environment at 28, 73, and 142 GHz \cite{Rap13a,Rappaport_2015,Mac14a,MacCartney_2019,Xing21a}, and four measurement campaigns were conducted in four different factory environments at 142 GHz \cite{Ju_2022,Ju_2024a}.

\begin{figure}[]
	\centering
	\includegraphics[width=0.35\textwidth]{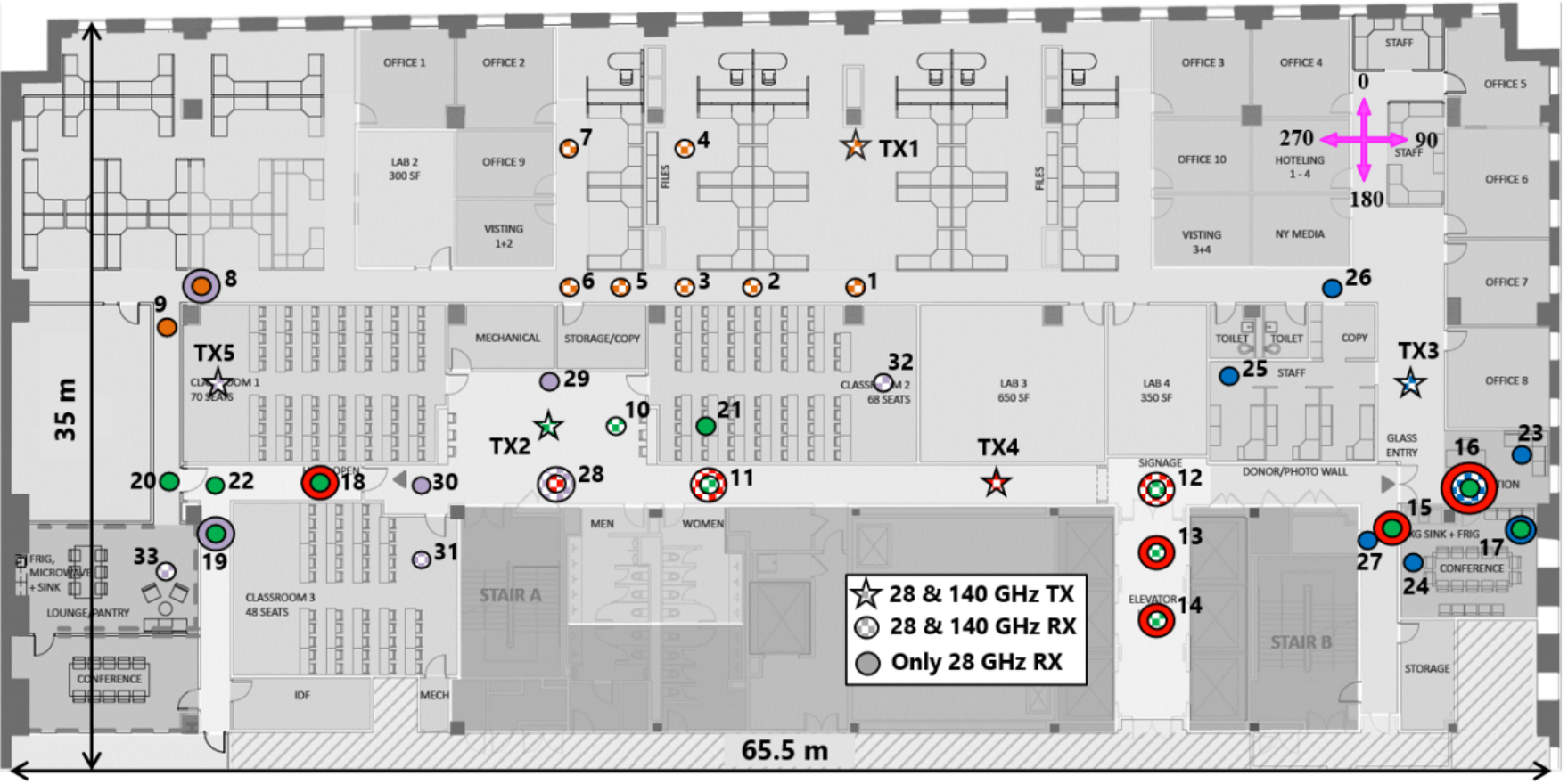}
	\caption{ Map of the indoor office environment displaying the TX and RX locations where channel measurements were conducted at 28 and 142 GHz. 33 TX-RX location combinations were measured at 28 GHz and 22 TX-RX combinations were measured at 142 GHz \cite{Mac15b,Ju_2021}. The TX locations are depicted by stars, the RX locations are depicted by circles. RX locations paired with a TX location are denoted in the same color. The links measured at both frequencies are depicted by a checkerboard texture, while links only measured at 28 GHz are depicted by a solid texture \cite{Mac15b,Ju_2021}.}
	\label{fig:Indoor_locations}
\end{figure}

Indoor office measurement campaigns were conducted at 28 and 142 GHz in an indoor open-office on the 9th floor of 2 Metrotech Center in 2014 and 2019, respectively \cite{Mac15b,Ju_2021}. The environment consisted of cubicles, private offices, classrooms, glass doors and elevators. The walls were made of drywall with metallic frames and metal studs interspersed. A map of the TX and RX measurement locations for indoor scenario is shown in Fig. \ref{fig:Indoor_locations}. 

In the indoor measurement campaigns at 28 and 142 GHz, the horn antennas were mounted on electronically steerable gimbals with sub-degree accuracy in the azimuth and elevation plane and rotated in steps of the antenna HPBW. The TX gimbal was at a height of 2.5 m, just below the ceiling (the ceiling was 2.75 m high) to replicate the location where mmWave base station could be deployed, while the RX gimbal was at a height of 1.5 m, the typical user equipment (UE) height.

\begin{figure}
	\centering

    {\includegraphics[width=0.35\textwidth]{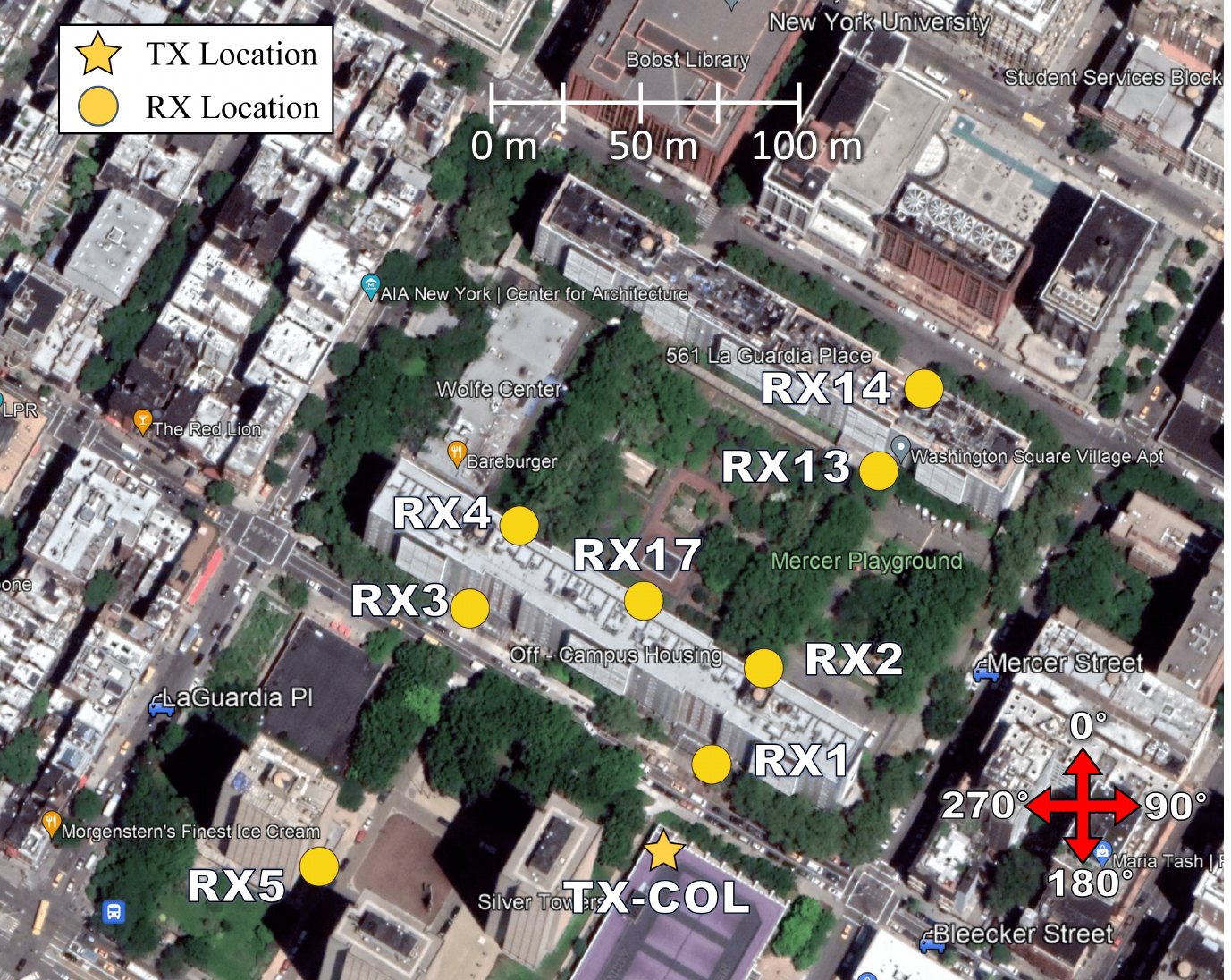}\label{fig:Outdoor_COL}}

    \vspace{0.2 in}
    \centering

    {\includegraphics[width=0.35\textwidth]{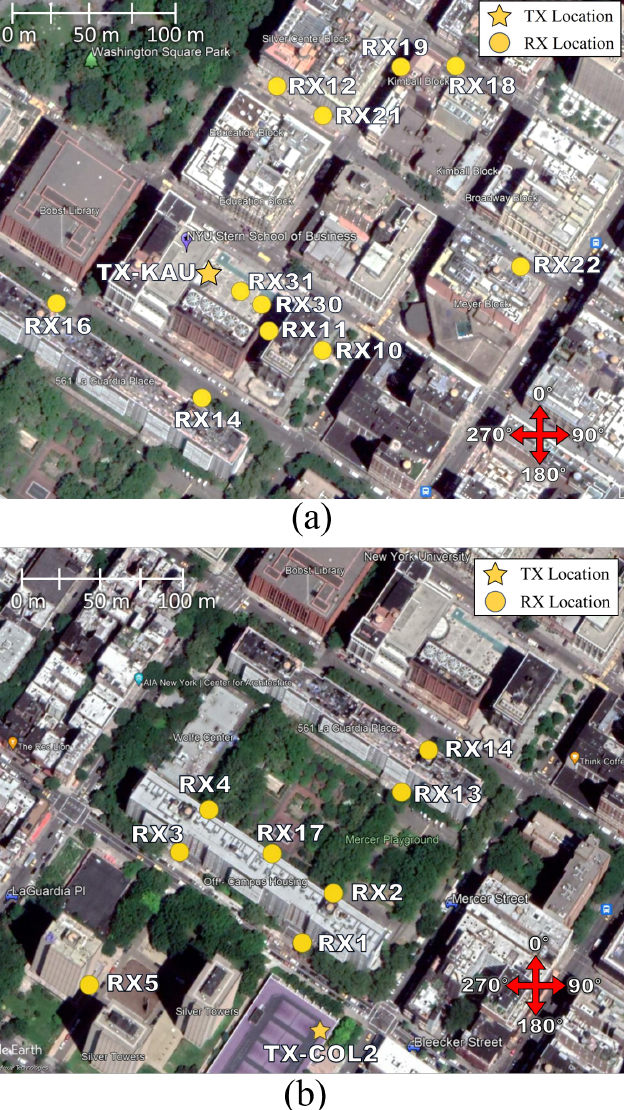}\label{fig:Outdoor_KAU}}
	
	\caption{ Map of the outdoor environment where the 28 GHz measurements were conducted, displaying the TX and RX locations \cite{Rappaport_2015}.}
	\label{fig:28GHz_Outdoor}
\end{figure}

\begin{figure}[]
	\centering
	\includegraphics[width=0.35\textwidth]{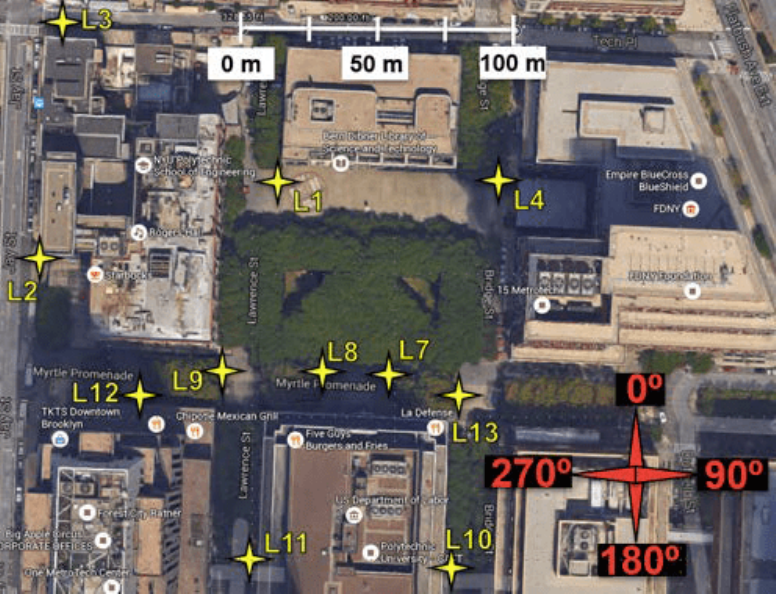}
	\caption{ Map of the outdoor O.S environment where the 73 GHz measurements were conducted, displaying the TX and RX locations \cite{Mac14a,MacCartney_2019}.}
	\label{fig:73GHz_Outdoor}
\end{figure}

\begin{figure}[]
	\centering
	\includegraphics[width=0.35\textwidth]{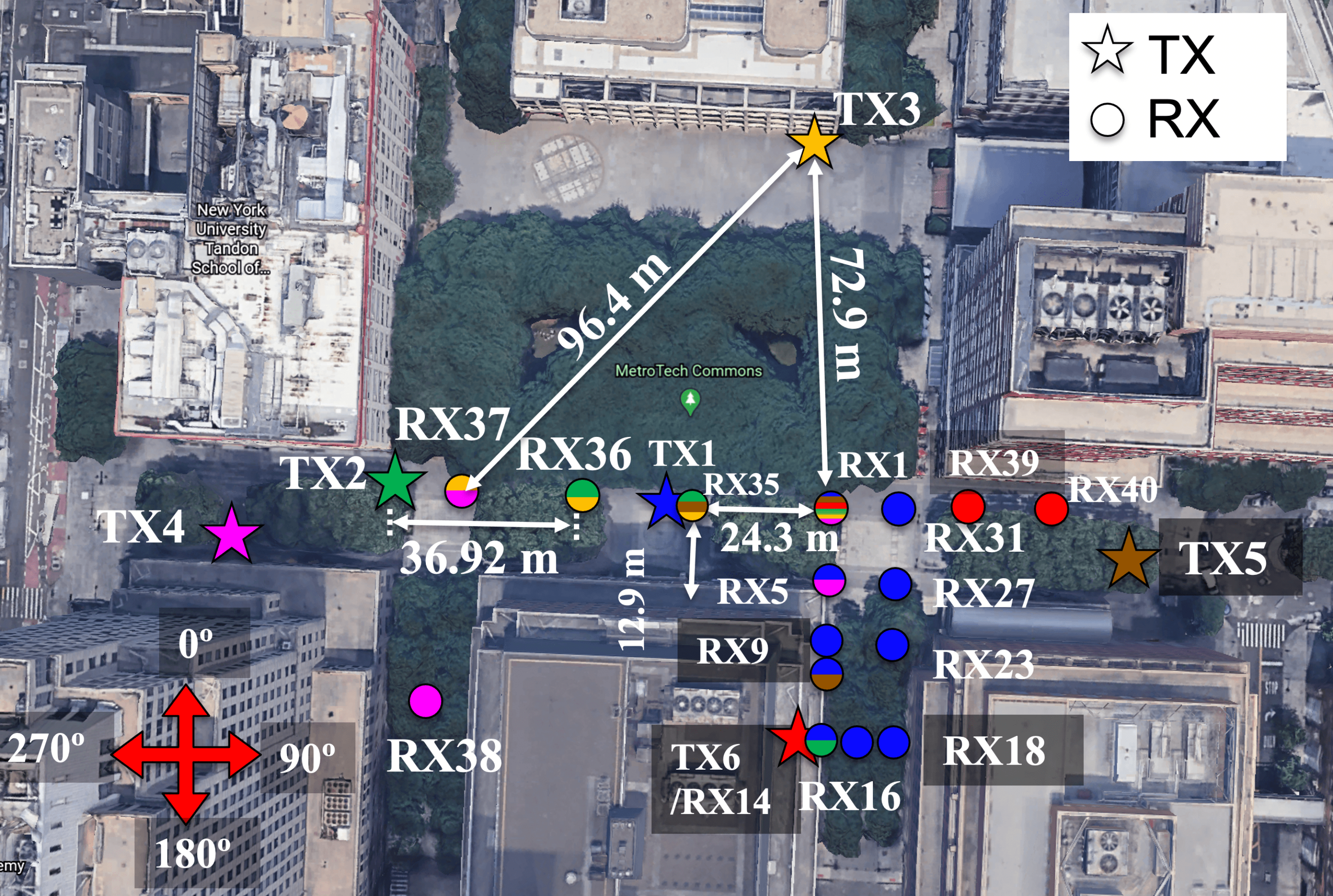}
	\caption{ Map of the outdoor O.S environment where the 142 GHz measurements were conducted, displaying the TX and RX locations. RX locations paired with a TX location are denoted in the same color \cite{Xing21a}.}
	\label{fig:140GHz_Outdoor}
\end{figure}

Outdoor measurement campaigns were conducted at 28, 73, and 142 GHz \cite{Rap13a,Rappaport_2015,Mac14a,MacCartney_2019,Xing21a}. The 28 GHz outdoor measurements were conducted at the NYU campus in downtown Manhattan, which was an urban environment surrounded by general university areas and high rise buildings \cite{Rappaport_2015}. The 73 and 142 GHz outdoor measurements were conducted in an open square (O.S.) environment on the NYU Tandon campus in downtown Brooklyn \cite{Mac14a,MacCartney_2019,Xing21a}. The TX-RX locations used in the 28 GHz measurement campaign are shown in Fig. \ref{fig:28GHz_Outdoor}, the TX-RX locations of the 73 GHz measurement campaign are shown in Fig. \ref{fig:73GHz_Outdoor}, while the locations used in the 142 GHz measurement campaign are shown in Fig. \ref{fig:140GHz_Outdoor}. Reflective metallic objects such as benches and lampposts were present in the environment. An orchard of cherry trees was in the center of the environment, with buildings that were four to fifteen stories surrounding the four sides of the orchard. 


In the outdoor measurement campaign at 28 GHz, the first TX was located on the rooftop of Coles Sports Center, at a height of 7 m. The second TX was on the fifth floor balcony of Kaufman Center, at a height of 17 m. All the RXs were at a height of 1.5 m, the typical mobile UE height. Similarly, in the outdoor measurement campaign at 73 GHz, the TX heights were 4 m above ground level, while the RX heights were 1.4 m.

\begin{figure}[]
	\centering
	\includegraphics[width=0.35\textwidth]{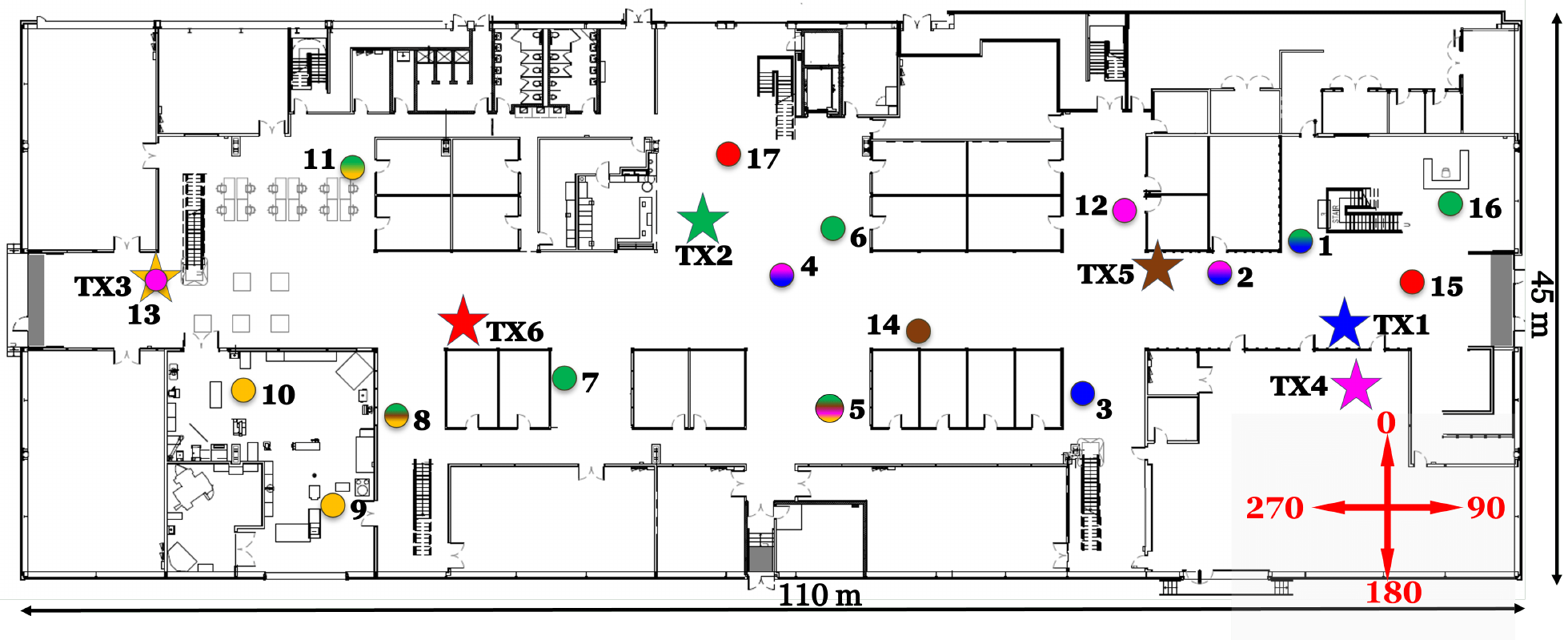}
	\caption{ Map of the floorplan of Factory A, depicting the TX-RX measurement locations at 142 GHz. RX locations paired with a TX location are denoted in the same color \cite{Ju_2024a}.}
	\label{fig:NEWLAB_locations}
\end{figure}

\begin{figure}[]
	\centering
	\includegraphics[width=0.35\textwidth]{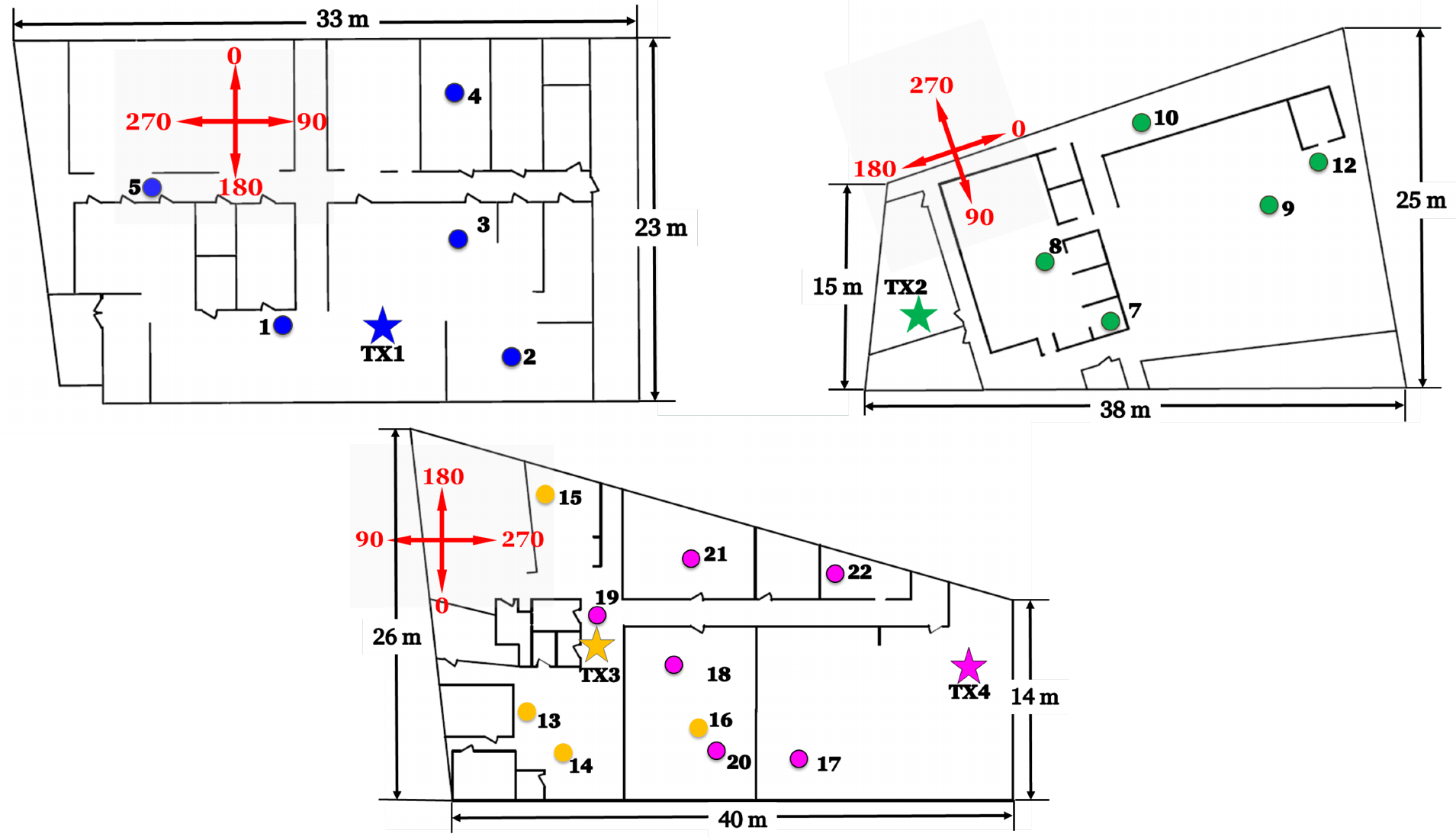}
	\caption{ Map of the floorplan of Factory B, depicting the TX-RX measurement locations at 142 GHz. RX locations paired with a TX location are denoted in the same color \cite{Ju_2024a}.}
	\label{fig:NEPTUNE_locations}
\end{figure}

\begin{figure}[]
	\centering
	\includegraphics[width=0.35\textwidth]{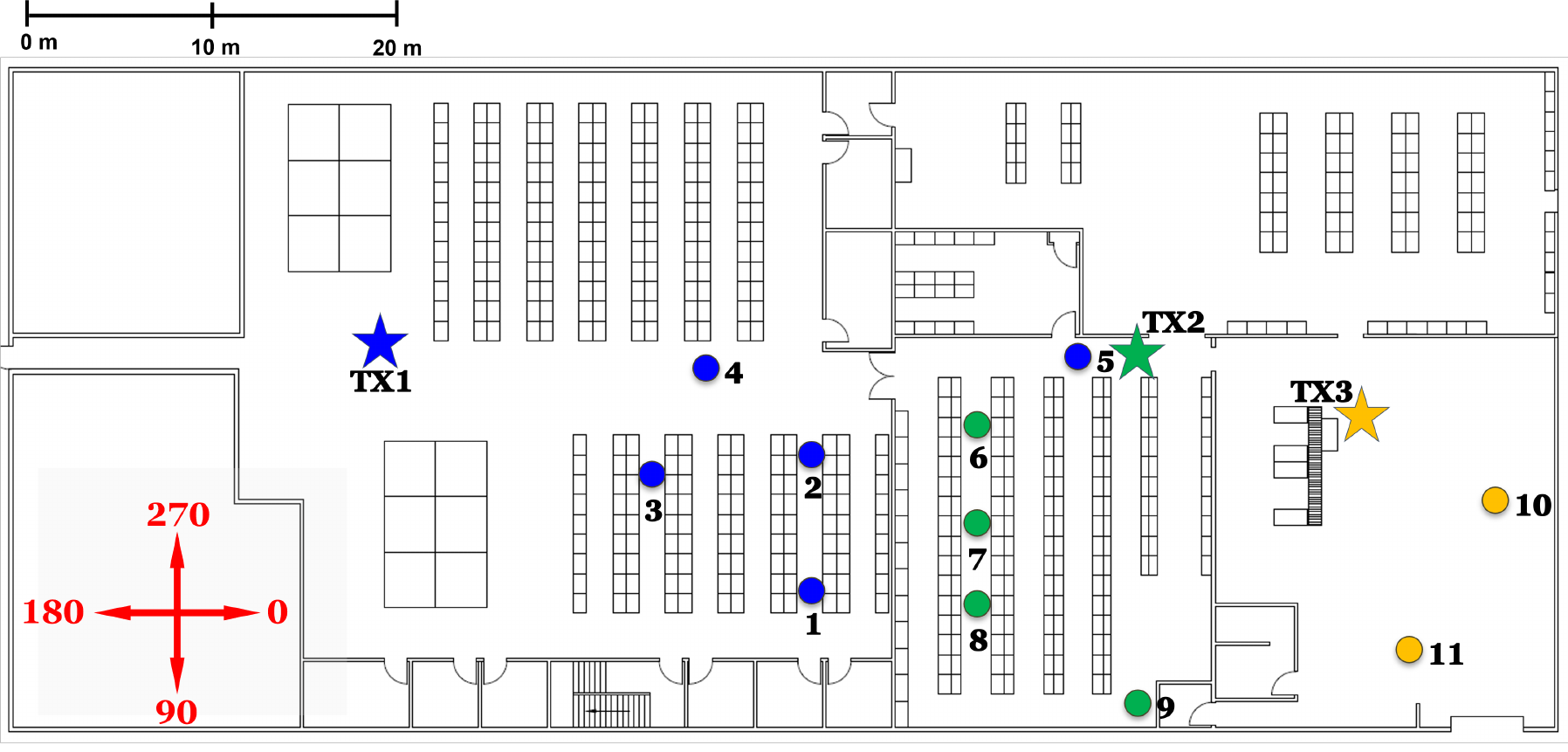}
	\caption{ Map of the floorplan of Factory C, depicting the TX-RX measurement locations at 142 GHz. RX locations paired with a TX location are denoted in the same color \cite{Ju_2024a}.}
	\label{fig:DEERPARK_locations}
\end{figure}

\begin{figure}[]
	\centering
	\includegraphics[width=0.35\textwidth]{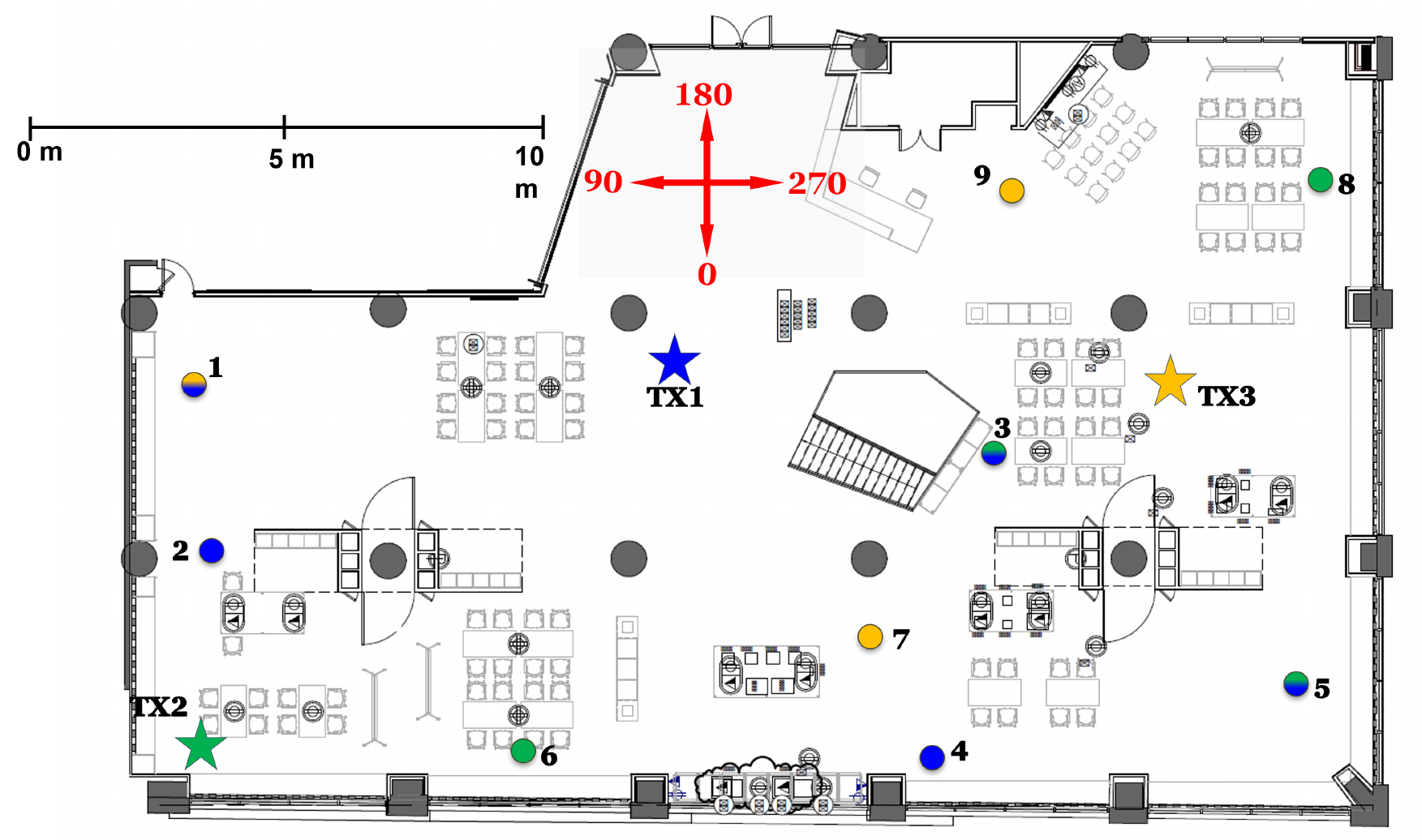}
	\caption{ Map of the floorplan of Factory D, depicting the TX-RX measurement locations at 142 GHz. RX locations paired with a TX location are denoted in the same color \cite{Ju_2024a}.}
	\label{fig:MAKERSPACE_locations}
\end{figure}

Four factory measurement campaigns were conducted at 142 GHz \cite{Ju_2022,Ju_2024a}. The first factory site was two-storied, large factory building in Brooklyn, NY where multiple tech prototyping companies shared office space, with the TX-RX locations depicted in Fig. \ref{fig:NEWLAB_locations}. The factory had shared open-seating areas, dedicated meeting rooms with glass panes and metallic frames, lockers, and metal elevators. The second factory was a a medium-sized electronics manufacturing facility in Brooklyn, NY. Numerous manufacturing machines were present in the environment, leading to a dense multipath environment. The TX-RX locations for Factory B are shown in Fig. \ref{fig:NEPTUNE_locations}. The third factory was a medium sized warehouse facility in Long Island, NY, depicted in Fig. \ref{fig:DEERPARK_locations}. The warehouse had several metallic storage aisles stocked with cardboard boxes, and a package inspection room consisting of multiple desks for package inspection, along with a conveyor belt. The fourth factory as illustrated in Fig. \ref{fig:MAKERSPACE_locations} was a single room manufacturing space in Brooklyn, NY. Prototyping machines such as 3D printing machines and PCB production equipment were present in the environment. 

\section{Utilizing NYURay to Analyze Measured Data}
A ray tracer accurately simulates site-specific real-world wireless channels, to provide AoA, AoD, ToF and received signal strength information of arriving multipath components. The mmWave and sub-THz channel measurements described in Section \ref{sec:measurements} utilized a sliding-correlation based channel sounder, with the TX and RX separated by several tens of meters, with no cable connecting the TX and RX (cables over such long distances could pose trip hazards). In the indoor and outdoor measurement locations, GPS receivers were unreliable \cite{Mac17a}, due to which the TX and the RX were not synchronized. Although rubidium clocks were used in some of the measurement campaigns (in the sub-THz factory measurements \cite{Ju_2024a} and the 73 GHz outdoor measurements \cite{Mac14a,MacCartney_2019}), other measurement campaigns lacked the means to obtain the absolute time of arrival of multipath components, which was needed for aligning directional PDP to create omnidirectional PDP \cite{Sun_2015}. In such environments, the absolute time of arrival of multipath components predicted by NYURay was used for wireless channel analysis \cite{Shakya_2023}. 

NYURay was also a vital tool for processing the multipath data captured in the mmWave and sub-THz measurement campaigns, to make sure we could identify and crosscheck the AoA and AoD of measured multipath components from the field data, which sometimes had logging errors (e.g. due to one-off data recording error by the channel sounder software) and human errors (e.g. incorrectly recording the direction the TX and RX were pointing) \cite{Shakya_2023, Ju_2021}.

\section{Calibrating NYURay to Real-World Measurements}\label{sec:calibration}

NYURay was calibrated using directional channel propagation measurements from the indoor, outdoor and factory environments at mmWave and sub-THz frequencies, based on measurements reported in \cite{Ju_2021,Mac15b,Rap13a,Rappaport_2015,Mac14a,MacCartney_2019,Xing_2018,Xing21a,Ju_2022b, Shakya_2023}. The power of individual directional multipath components was calibrated by analyzing the directional propagation loss equations \cite{Kanhere_2023}. Calibration was performed on a system with an Intel i7-3770 CPU with 16 GB of RAM. With scattering disabled, the average ray tracing computation for each location took 93 ms. With scattering enabled, the average computation time increased to 1.1 s per location.

In this manuscript, we optimize the reflection and penetration characteristics of various materials, while keeping the constant scattering parameters constant for all materials, with $ \Lambda = 0.8, \alpha_i=\alpha_R=10 $, and $ S $ = 0.1, where S is the scattering parameter related to $ E_{so} $ as described in \cite{Degli_Eposti2007}. 

Optimization of the reflection and penetration losses may be done in the linear domain (mW) or in the decibels domain (dB). In this article, the reflection and penetration losses were optimized in the decibel domain using a closed-form linear least square optimizer. A closed form solution is unknown for optimizing the reflection and penetration losses in the linear domain. Due to the high dimensionality of the optimization problem, iterative optimizers such as gradient descent converged to local minimal (sub-optimal) solutions when the initial values of reflection and penetration loss parameters were not close to optimal parameter values.

\subsection{Optimization in the Decibel Domain}
Consider a single multipath component $ j $ arriving at the RX after traversing a distance of $ d $ m, propagating via specular reflections and/or penetration. NYURay models the power carried by a specular multipath component($ P_{j,R} $) as follows:

\begin{align}
&	P_{j,R} [dBm] = P_{j,TX} [dBm] +G_{j,T} [dBi] +G_{j,R} [dBi] \nonumber\\
&	-FSPL(d_j,f) [dB] -\sum_{i=1}^{N}\left(w^{pen}_{i,j}L^{pen}_{i,j}+ w^{ref}_{i,j}L^{ref}_{i,j}\right) [dB], \label{eq:pl}
\end{align}
where $ P_{j,TX} $ is the RF power at the TX in dBm, $  G_{j,T} $ and $ G_{j,R} $ are the antenna gains (in dBi) in the direction of departure and arrival of the signal at the TX and RX respectively, $ d_j $ is the distance traveled by the signal from the TX to the RX in meters, $ FSPL(d_j,f) $ is the free space path loss (in dB) at a distance $ d_j $ and a frequency $ f $ (GHz). $ N $ is the number of material types in the environment. The reflection loss and partition loss of each material will be estimated via ray tracer calibration. $w^{pen}_{i,j} L^{pen}_{i} $ is the total power lost (in dB) due to penetration loss of the multipath component $ j $ with obstructions of material type $ i $. The weight $ w^{pen}_{i,j} $ is set to be equal the number of times the multipath penetrates through obstacles with material type $ i $, while $ L^{pen}_{i} $ is the power lost at each such penetration. $w^{ref}_{i,j} L^{ref}_{i} $ is the total power lost (in dB) due to reflection of the multipath component $ j $ with obstructions of material type $ i $. The weight $ w^{ref}_{i,j} $ is set to be equal the number of times the multipath reflects off obstacles with material type $ i $, while $ L^{ref}_{i} $ is the power lost at each such reflection.

Directional real-world measurements capture the power contained in one or more multipath components. Due to the directional horn antennas used during channel measurements, typically one strong multipath component was observed. The power measured in the strongest multipath component was compared to the power predicted via \eqref{eq:pl}.

If the power contained in the strongest multipath component (in decibels) is given by $ P_{j,meas} $, the following least-squares objective function ($OF_{dB}$) must be minimized over all directional measurements:
\begin{align}
	OF_{dB}=	&\sum_{j=1}^{M} \left( P_{j,meas}-P_{j,R}\right)^2\nonumber\\
	&=\sum_{j=1}^{M}\Big(P_{j,TX} +G_{j,T} +G_{j,R} -FSPL(d_j,f)\nonumber\\
	&-\sum_{i=1}^{N}\left(w^{pen}_{i,j}L^{pen}_{i}+ w^{ref}_{i,j}L^{ref}_{i}\right)-P_{j,meas}\Big)^2 \label{eq:LS}
\end{align}
where all quantities are in log-scale, $ L $ is the squared loss function that is to be minimized and $ M $ is the number of multipath components over which calibration is done. A correspondence between the multipath component predicted by the ray tracer and the measured multipath component was established by choosing the simulated multipath component with the least deviation in angle of arrival and angle of departure from the measured angles of arrival and departure.

Consider each individual summation term in \eqref{eq:LS}. The TX power of the directional measurement was obtained using a power meter during the calibration of the channel sounder, as described in \cite{Xing_2018_VTC}. The antenna gains in the direction of departure and arrival of the multipath component were obtained by utilizing the antenna pattern of the TX and RX horn antennas, which were measured in the laboratory as described in Section \ref{sec:equipment}. The path length traversed by a particular multipath component, as predicted by NYURay, is independent of the electrical properties of the material in the environment (depending only on the geometry of the environment). Thus, the free space path loss of each multipath component ($ FSPL(d_j,f) $) was obtained by utilizing an uncalibrated version of the ray tracer, which provided accurate information of path length (although the power of the multipath component is not accurate when an uncalibrated ray tracer is used). To ensure that each multipath component was predicted by the uncalibrated ray tracer with sufficient power, the reflection coefficient of all materials was initialized to $\Gamma = 0.9$ (highly reflective) with a partition loss of 1 dB. The material properties are fine-tuned as part of the calibration process. Finally, $ P_{j,meas} $ is the power contained in the multipath component as obtained from the directional field measurement. Thus, all quantities other than $ \sum_{i=1}^{N}\left(w^{pen}_{i,j}L^{pen}_{i}+ w^{ref}_{i,j}L^{ref}_{i}\right) $ are known for the directional measurement corresponding to multipath component $ j $. Thus, \eqref{eq:LS} may be simplified to:

\begin{align}
	{OF_{dB}}&= \sum_{j=1}^{M}\left(A_j-\sum_{i=1}^{N}\left(w^{pen}_{i,j}L^{pen}_{i}+ w^{ref}_{i,j}L^{ref}_{i}\right)\right)^2  \\
	&= \sum_{j=1}^{M}\left(A_j-\mathbf{w_j}\mathbf{L}\right)^2  \label{eq:loss_log}
\end{align}
where $ \mathbf{w_j} = [w^{pen}_{1,j} w^{pen}_{2,j} \dotsc w^{pen}_{N,j} w^{ref}_{1,j} w^{ref}_{2,j} \dotsc w^{ref}_{N,j} ]$ and $ \mathbf{L} = [L^{pen}_{1} L^{pen}_{2} \dotsc L^{pen}_{N} L^{ref}_{1} L^{ref}_{2} \dotsc L^{ref}_{N}]^T  $.

The optimal values of $ L^{pen}_{i} $and $ L^{ref}_{i} $ ($\hat{L}$) are given by the least squares solution to the following set of linear equations:
\begin{align}
	\mathbf{W} \mathbf{L} = \mathbf{A},
\end{align}
where 
\begin{align}
	\mathbf{W} = \begin{bmatrix}
		\mathbf{w_1} \\
		\mathbf{	 w_2} \\ 
		\vdots \\
		\mathbf{  w_N}
	\end{bmatrix}
\end{align}
Thus, $ \hat{L} $ is given by
\begin{align}
	\hat{L} = (\mathbf{W}^T\mathbf{W})^{-1} \mathbf{W}^T \mathbf{A}
\end{align}

\subsection{Optimization in the Linear Domain}

In the previous section, the reflection and penetration losses of building materials were determined via optimization in the log-domain, which is more convenient due to the existence of a closed form solution. However, if optimization is done in the linear domain, the Objective Function ($OF_{lin}$) may be expressed as:
\begin{align}
	{OF_{lin}}&= \sqrt{\left(\frac{\sum_{j=1}^{M}\left(B_j-\prod_{i=1}^{N}{l^{pen}_{i}}^{w^{pen}_{i,j}} {l^{ref}_{i}}^{w^{ref}_{i,j}}\right)}{M}^2 \right)} \label{eq:loss_linear}
\end{align}
where $ l^{pen}_{i} $ and $l^{ref}_{i}$ are the (linear) penetration and reflection coefficients of material $ i $ respectively and $ B_j = \dfrac{p_{j,TX}g_{j,T}g_{j,R} }{(FSPL_{lin}(d_j,f)p_{j,meas})}$. Here, $ g_{j,T} $ and $ g_{j,R} $ are the antenna gains (linear) in the direction of departure and arrival of the signal at the TX and RX respectively. $ FSPL_{lin}(d_j,f) $ is the free space path loss factor (linear) at a distance $ d_j $ and frequency $ f $ (GHz), $ p_{j,TX} $ is the RF power at the TX in mW and $ p_{j,meas} $ is the RF power measured at the RX in mW.

\begin{figure}[]
	\centering
	\includegraphics[width=0.5\textwidth]{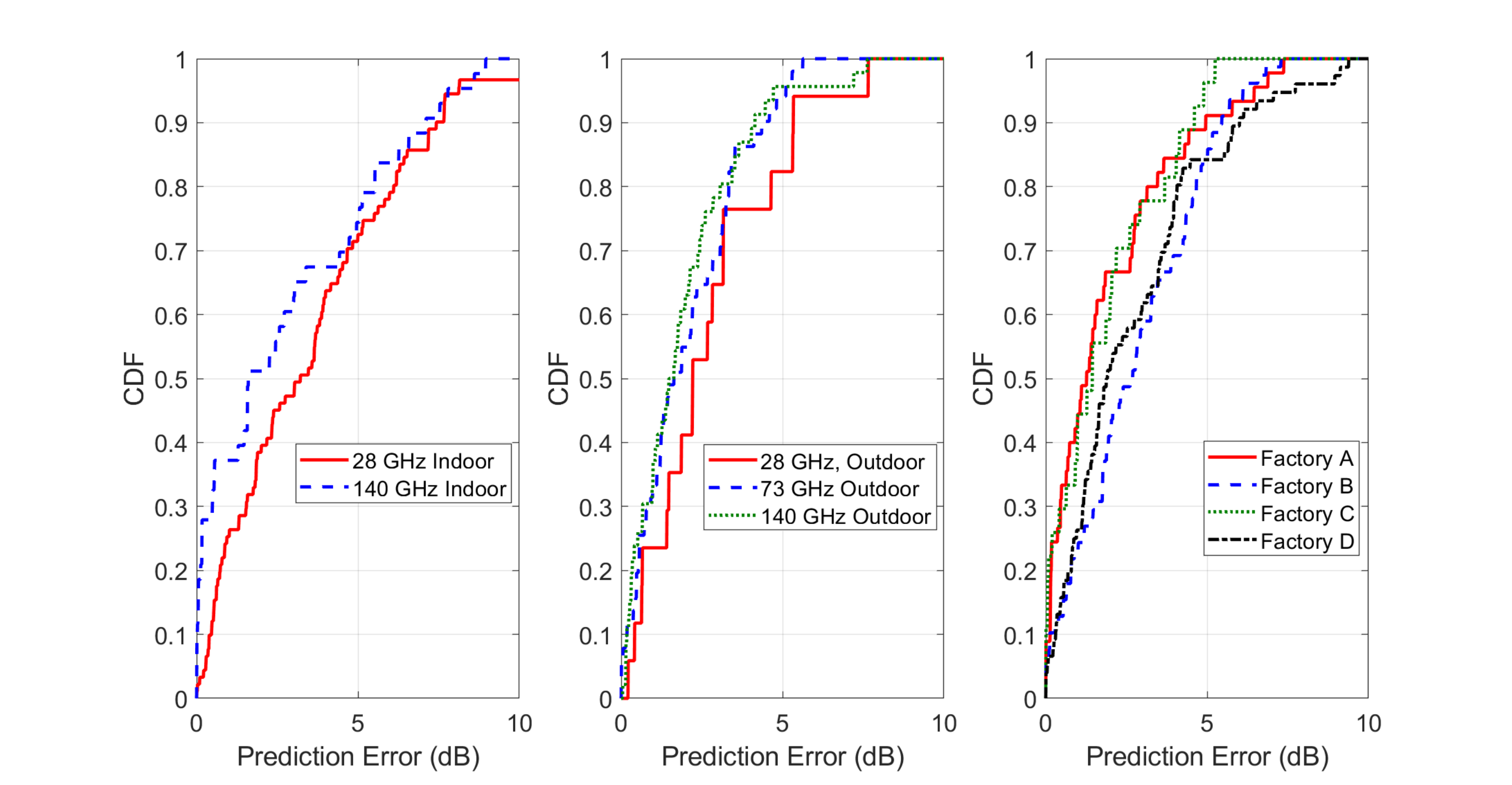}
	\caption{The absolute error in the multipath power predicted by NYURay ( $ |L| $ ) compared to the measured power, over the indoor, outdoor, and factory measurement campaigns \cite{Ju_2021,Mac15b,Rap13a,Rappaport_2015,Mac14a,MacCartney_2019,Xing_2018,Xing21a,Ju_2022b}.}
	\label{fig:power_error}
\end{figure}

\begin{figure}[]
	\centering
	\includegraphics[width=0.35\textwidth]{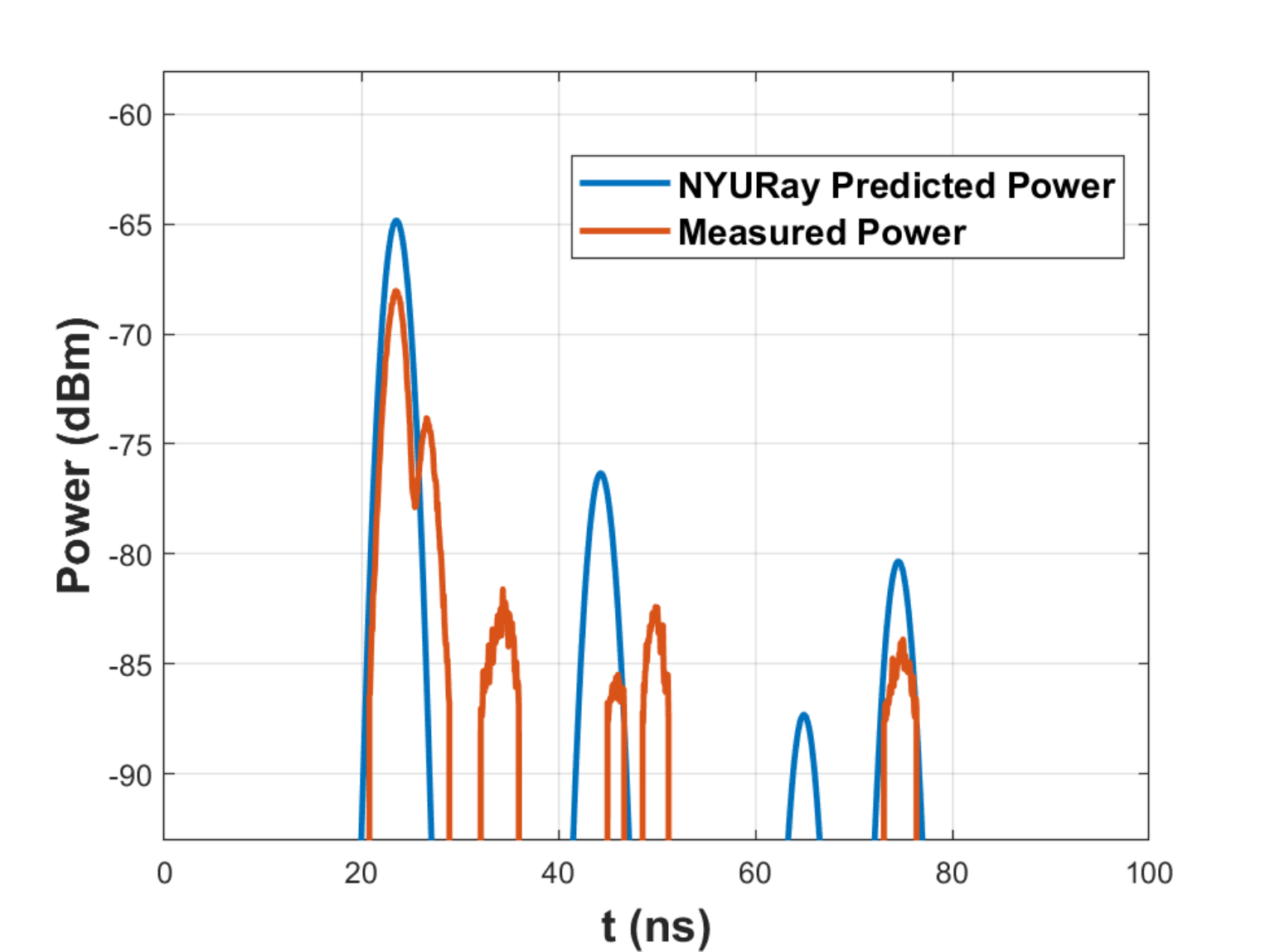}
	\caption{Comparison between the measured and simulated PDP at an indoor location at 28 GHz after calibration \cite{Mac15b,Kanhere_2023}}
	\label{fig:directional_pdp}
\end{figure}

Similarly, in the presence of non-negligible diffraction and scattering, a closed form solution for the interaction losses of materials is unknown. Hence, iterative approaches such as simulated annealing can be utilized to obtain the diffraction and scattering parameters \cite{Gan_2014}. In simulated annealing, a solution is iteratively improved by exploring neighboring solutions, with a non-zero probability of accepting worse neighboring solutions in each iteration step. This approach reduces the likelihood of the algorithm getting stuck in local optima, thereby enhancing the chances of finding a global optimum.

\section{ Performance of NYURay for Predicting mmWave and sub-THz Measurements }
To evaluate the performance of NYURay, the measured directional multipath power (in dB) was compared with the power predicted by NYURay for the nine measurement campaigns. A comparison of the value of the loss functions introduced in \eqref{eq:loss_log} and \eqref{eq:loss_linear} via linear and logarithmic optimization techniques is provided in Table \ref{tbl:lin_vs_dB}. The loss function in the decibel domain ($ OF_{dB} $) is better optimized via the log optimization technique, while the loss function in the linear domain ($ OF_{Lin} $) is better optimized by the linear optimization technique.

\begin{table*}[]
	\caption{\centering Performance comparison of the linear and logarithmic optimization techniques in indoor, outdoor, and factory environments at 28, 73, and 140 GHz using measurement data reported in \cite{Ju_2021,Mac15b,Rap13a,Rappaport_2015,Mac14a,MacCartney_2019,Xing_2018,Xing21a,Ju_2022b}.}
	\centering
	\small
	\begin{tabular}{|c|c|c|c|c|}
		\hline
		Measurement Campaign & \multicolumn{2}{c|}{$ OF_{dB} $} & \multicolumn{2}{c|}{$ OF_{Lin} $} \\\hline
		& Log. Opt.    & Lin. Opt.   & Log. Opt.    & Lin. Opt.    \\\hline
		28 GHz Indoor        & 4.4         & 5.4        & 0.16        & 0.13        \\\hline
		140 GHz Indoor       & 3.9         & 5.7        & 0.11        & 0.09        \\\hline
		28 GHz Outdoor       & 3.2         & 3.5        & 0.03        & 0.02        \\\hline
		73 GHz Outdoor       & 2.6         & 3.3        & 0.05        & 0.04        \\\hline
		140 GHz Outdoor      & 2.6         & 3.0        & 0.11        & 0.05        \\\hline
		140 GHz Factory A    & 2.3         & 2.4        & 0.06        & 0.06        \\\hline
		140 GHz Factory B    & 3.4         & 3.8        & 0.10        & 0.09        \\\hline
		140 GHz Factory C    & 2.5         & 5.6        & 0.08        & 0.05        \\\hline
		140 GHz Factory D    & 3.6         & 4.1           & 0.13            &      0.12      \\\hline
	\end{tabular}
	
	\label{tbl:lin_vs_dB}
\end{table*}

\par The performance of NYURay was assessed by comparing the measured and simulated secondary channel statistics, particularly the RMS delay spread \cite{Rap02a} and the RMS angular spread \cite{3GPP.38.901}, in addition to directional multipath power (which was directly calibrated by least squares estimate).

\begin{figure}[]
	\centering
	\includegraphics[width=0.5\textwidth]{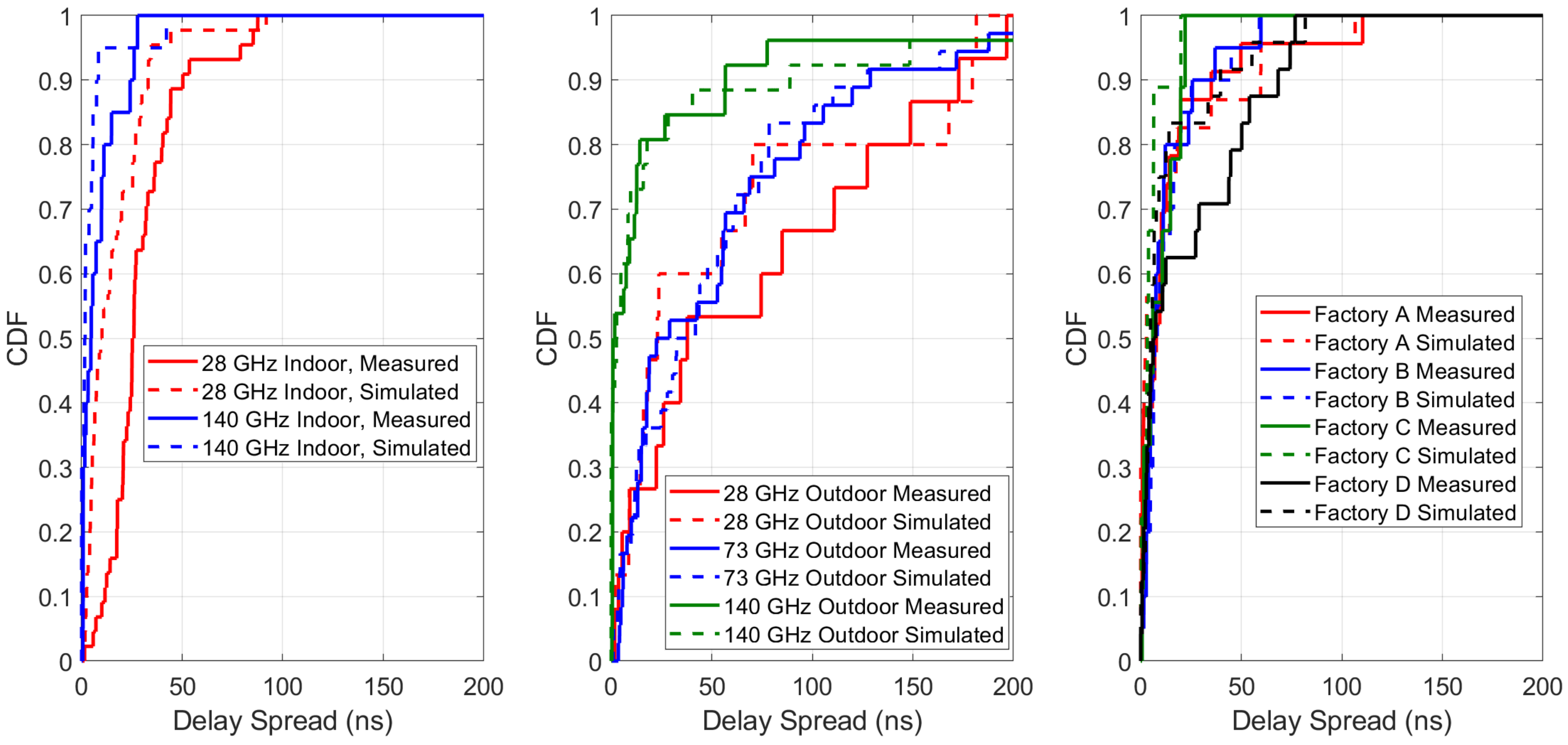}
	\caption{Comparison between the measured and simulated RMS delay spread for the indoor, outdoor, and factory measurement campaigns \cite{Ju_2021,Mac15b,Rap13a,Rappaport_2015,Mac14a,MacCartney_2019,Xing_2018,Xing21a,Ju_2022b}.}
	\label{fig:delay_spread}
\end{figure}

The RMS delay spread is a measure of the duration over which power is received in a multipath wireless channel \cite{Rap02a}. As seen in Fig. \ref{fig:delay_spread}, good agreement was observed between the measured RMS delay spread and the NYURay predicted RMS delay spread. The RMS delay spread at mmWave frequencies was observed to be greater than the RMS delay spread at sub-THz frequencies, with the greatest spread observed at 28 GHz in the outdoor environment. A reduction in RMS delay spread was observed at higher frequencies due to greater penetration loss through various obstructions in the environment, leading to fewer (and weaker) multipath components.

\begin{figure}[]
	\centering
	\includegraphics[width=0.5\textwidth]{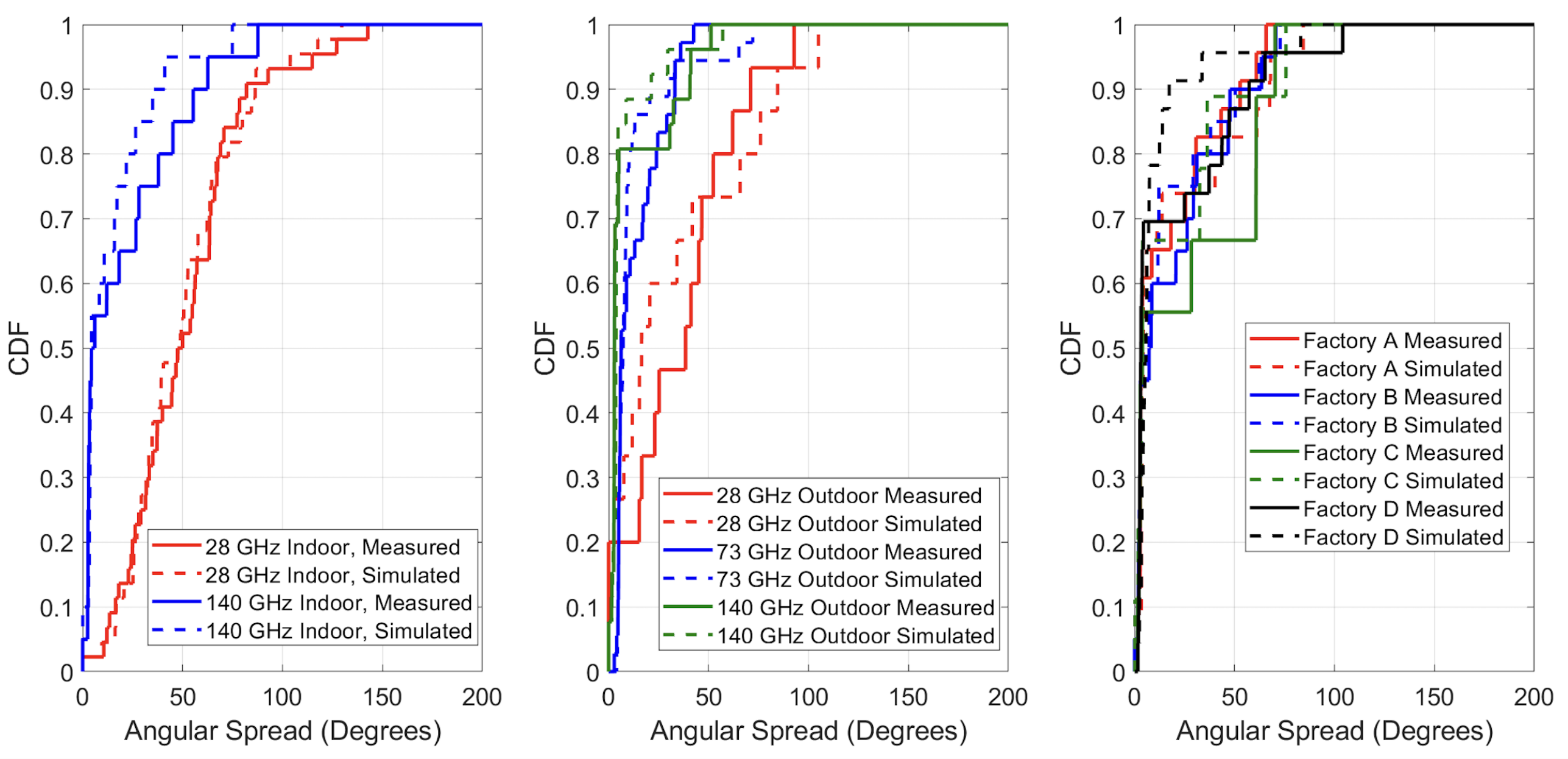}
	\caption{Comparison between the measured and simulated RMS angular spread for the indoor, outdoor, and factory measurement campaigns \cite{Ju_2021,Mac15b,Rap13a,Rappaport_2015,Mac14a,MacCartney_2019,Xing_2018,Xing21a,Ju_2022b}.}
	\label{fig:angular_spread}
\end{figure}

The RMS angular spread is a measure of dispersion of power over arriving incident angles \cite{3GPP.38.901}. As seen in Fig. \ref{fig:angular_spread}, there is a good agreement between the measured and simulated RMS angular spread.

A comparison between the RMS delay spread and RMS angular spread as predicted by NYURay using the linear and decibel optimization techniques, measured in the indoor office, outdoor, and factory environments is provided in Tables \ref{tbl:stats_indoor_AS}, \ref{tbl:stats_indoor_DS}, \ref{tbl:stats_outdoor_AS}, \ref{tbl:stats_outdoor_DS}, \ref{tbl:stats_Factory_AS}, and  \ref{tbl:stats_Factory_DS}. As is seen from the difference in the measured and predicted mean RMS angular spread and RMS delay spreads in the indoor office and outdoor environments, NYURay under-predicted the angular and delay spreads by 10\% and 20\% respectively, in indoor and outdoor environments. The underprediction was due to the incomplete environmental knowledge available for ray tracing - small objects in the environment resulted in signal reflections that were not predicted by the ray tracer. As seen in the tables, the accuracy of both optimization techniques was similar. Hence the optimization technique in the decibel domain is preferred due to the closed form nature and guaranteed convergence.

\renewcommand{\arraystretch}{1.2}
\begin{table*}[]
	\caption{\centering Difference ($\Delta$) between the RMS angular spread (Meas.) using channel measurements and predicted (Pred.) by NYURay in the indoor office environment at 28 GHz and 142 GHz using measurement data reported in \cite{Mac15b, Ju_2021}.}
	\centering
	\small
	\begin{tabular}{|c|ccccc|ccccc|}
		\hline
		\begin{tabular}[c]{@{}c@{}}Measurement\\ Campaign\end{tabular} & \multicolumn{5}{c|}{RMS angular spread ($ \mu $, degrees)}                                                                                                                                                                                                                                                                                           & \multicolumn{5}{c|}{RMS angular spread ($ \sigma $, degrees)}                                                                                                                                                                                                                                                                                            \\ \hline
		& \multicolumn{1}{c|}{Meas.} & \multicolumn{1}{c|}{\begin{tabular}[c]{@{}c@{}}Pred. \\ (log. Opt.)\end{tabular}} & \multicolumn{1}{c|}{\begin{tabular}[c]{@{}c@{}}$ \Delta $ \\ (log Opt.)\end{tabular}} & \multicolumn{1}{c|}{\begin{tabular}[c]{@{}c@{}}Pred. \\ (lin. Opt.)\end{tabular}} & \begin{tabular}[c]{@{}c@{}}$ \Delta $ \\ (lin. Opt.)\end{tabular} & \multicolumn{1}{c|}{Meas.} & \multicolumn{1}{c|}{\begin{tabular}[c]{@{}c@{}}Pred. \\ (log. Opt.)\end{tabular}} & \multicolumn{1}{c|}{\begin{tabular}[c]{@{}c@{}}$ \Delta $ \\ (log Opt.)\end{tabular}} & \multicolumn{1}{c|}{\begin{tabular}[c]{@{}c@{}}Pred. \\ (lin. Opt.)\end{tabular}} & \begin{tabular}[c]{@{}c@{}}$ \Delta $ \\ (lin. Opt.)\end{tabular} \\ \hline
		28 GHz Indoor                                                  & \multicolumn{1}{c|}{51.5} & \multicolumn{1}{c|}{49.9}                                                        & \multicolumn{1}{c|}{1.6}                                                         & \multicolumn{1}{c|}{50.0}                                                            &       \multicolumn{1}{c|}{1.5}                                                       & \multicolumn{1}{c|}{30.2} & \multicolumn{1}{c|}{28.3}                                                        & \multicolumn{1}{c|}{0.9}                                                         & \multicolumn{1}{c|}{30.6}                                                            &      0.4                                                        \\ \hline
		140 GHz Indoor                                                 & \multicolumn{1}{c|}{20.3} & \multicolumn{1}{c|}{14.3}                                                        & \multicolumn{1}{c|}{6.0}                                                         & \multicolumn{1}{c|}{23.7}                                                            &                              3.4                                & \multicolumn{1}{c|}{25.0} & \multicolumn{1}{c|}{18.5}                                                        & \multicolumn{1}{c|}{6.5}                                                         & \multicolumn{1}{c|}{34.5}                                                            &      9.5                                                        \\ \hline
	\end{tabular}
	\label{tbl:stats_indoor_AS}
\end{table*}
\renewcommand{\arraystretch}{1.2}
\begin{table*}[]
	\caption{\centering Difference ($\Delta$) between the RMS delay spread measured (Meas.) using channel measurements and predicted (Pred.) by NYURay in the indoor office environment at 28 GHz and 142 GHz using measurement data reported in \cite{Mac15b, Ju_2021}.}
	\centering
	\small
	\begin{tabular}{|c|ccccc|ccccc|}
		\hline
		\begin{tabular}[c]{@{}c@{}}Measurement\\ Campaign\end{tabular} & \multicolumn{5}{c|}{RMS delay spread ($ \mu $, ns)}   & \multicolumn{5}{c|}{RMS delay spread ($ \sigma $, ns)}                                                                                                                                                                                                                                                                                            \\ \hline
		& \multicolumn{1}{c|}{Meas.} & \multicolumn{1}{c|}{\begin{tabular}[c]{@{}c@{}}Pred. \\ (log. Opt.)\end{tabular}} & \multicolumn{1}{c|}{\begin{tabular}[c]{@{}c@{}}$\Delta$ \\ (log Opt.)\end{tabular}} & \multicolumn{1}{c|}{\begin{tabular}[c]{@{}c@{}}Pred. \\ (lin. Opt.)\end{tabular}} & \begin{tabular}[c]{@{}c@{}}$\Delta$ \\ (lin. Opt.)\end{tabular} & \multicolumn{1}{c|}{Meas.} & \multicolumn{1}{c|}{\begin{tabular}[c]{@{}c@{}}Pred. \\ (log. Opt.)\end{tabular}} & \multicolumn{1}{c|}{\begin{tabular}[c]{@{}c@{}}$\Delta$ \\ (log Opt.)\end{tabular}} & \multicolumn{1}{c|}{\begin{tabular}[c]{@{}c@{}}Pred. \\ (lin. Opt.)\end{tabular}} & \begin{tabular}[c]{@{}c@{}}$\Delta$ \\ (lin. Opt.)\end{tabular} \\ \hline
		28 GHz Indoor                                                  & \multicolumn{1}{c|}{29.8} & \multicolumn{1}{c|}{15.7}                                                        & \multicolumn{1}{c|}{14.1}                                                         & \multicolumn{1}{c|}{15.0}                                                            &      14.8                                                        & \multicolumn{1}{c|}{18.7} & \multicolumn{1}{c|}{16.3}                                                        & \multicolumn{1}{c|}{2.4}                                                         & \multicolumn{1}{c|}{14.3}                                                            &       4.4                                                       \\ \hline
		140 GHz Indoor                                                 & \multicolumn{1}{c|}{7.9} & \multicolumn{1}{c|}{4.4}                                                        & \multicolumn{1}{c|}{3.5}                                                         & \multicolumn{1}{c|}{6.7}                                                            &       1.2                                                       & \multicolumn{1}{c|}{8.8} & \multicolumn{1}{c|}{9.3}                                                        & \multicolumn{1}{c|}{-0.5}                                                         & \multicolumn{1}{c|}{9.1}                                                            &      0.3                                                        \\ \hline
	\end{tabular}
	\label{tbl:stats_indoor_DS}
\end{table*}

\renewcommand{\arraystretch}{1.2}
\begin{table*}[]
	\caption{\centering Difference ($\Delta$) between the RMS angular spread measured (Meas.) using channel measurements and predicted (Pred.) by NYURay in the outdoor environments at 28 GHz, 73 GHz, and 142 GHz using measurement data reported in \cite{Rap13a,Rappaport_2015,Xing21a, Shakya_2023}.}
	\centering
	\small
	\begin{tabular}{|c|ccccc|ccccc|}
		\hline
		\begin{tabular}[c]{@{}c@{}}Measurement\\ Campaign\end{tabular} & \multicolumn{5}{c|}{RMS angular spread ($ \mu $, degrees)} & \multicolumn{5}{c|}{RMS angular spread ($ \sigma $, degrees)} \\ \hline
		& \multicolumn{1}{c|}{Meas.} & \multicolumn{1}{c|}{\begin{tabular}[c]{@{}c@{}}Pred. \\ (log. Opt.)\end{tabular}} & \multicolumn{1}{c|}{\begin{tabular}[c]{@{}c@{}}$\Delta$ \\ (log Opt.)\end{tabular}} & \multicolumn{1}{c|}{\begin{tabular}[c]{@{}c@{}}Pred. \\ (lin. Opt.)\end{tabular}} & \begin{tabular}[c]{@{}c@{}}$\Delta$ \\ (lin. Opt.)\end{tabular} & \multicolumn{1}{c|}{Meas.} & \multicolumn{1}{c|}{\begin{tabular}[c]{@{}c@{}}Pred. \\ (log. Opt.)\end{tabular}} & \multicolumn{1}{c|}{\begin{tabular}[c]{@{}c@{}}$\Delta$ \\ (log Opt.)\end{tabular}} & \multicolumn{1}{c|}{\begin{tabular}[c]{@{}c@{}}Pred. \\ (lin. Opt.)\end{tabular}} & \begin{tabular}[c]{@{}c@{}}$\Delta$ \\ (lin. Opt.)\end{tabular} \\ \hline
		28 GHz Outdoor & \multicolumn{1}{c|}{35.3} & \multicolumn{1}{c|}{32.1} & \multicolumn{1}{c|}{3.2} & \multicolumn{1}{c|}{32.2} & 3.1 & \multicolumn{1}{c|}{27.5} & \multicolumn{1}{c|}{34.7} & \multicolumn{1}{c|}{-7.2} & \multicolumn{1}{c|}{34.8} & -7.3 \\ \hline
		73 GHz Outdoor & \multicolumn{1}{c|}{13.3} & \multicolumn{1}{c|}{12.2} & \multicolumn{1}{c|}{1.1} & \multicolumn{1}{c|}{12.5} &0.8 & \multicolumn{1}{c|}{11.3} & \multicolumn{1}{c|}{15.3} & \multicolumn{1}{c|}{-4.0} & \multicolumn{1}{c|}{16.7} & -5.4\\ \hline
		140 GHz Outdoor & \multicolumn{1}{c|}{9.7} & \multicolumn{1}{c|}{7.1} & \multicolumn{1}{c|}{2.6} & \multicolumn{1}{c|}{7.3} &2.4 & \multicolumn{1}{c|}{15.0} & \multicolumn{1}{c|}{12.0} & \multicolumn{1}{c|}{3.0} & \multicolumn{1}{c|}{11.7} & 3.3\\ \hline
	\end{tabular}
	\label{tbl:stats_outdoor_AS}
\end{table*}
\renewcommand{\arraystretch}{1.2}
\begin{table*}[]
	\caption{\centering Difference ($\Delta$) between the RMS delay spread measured (Meas.) using channel measurements and predicted (Pred.) by NYURay in the outdoor environments at 28 GHz, 73 GHz, and 142 GHz using measurement data reported in \cite{Rap13a,Rappaport_2015,Xing21a, Shakya_2023}.}
	\centering
	\small
	\begin{tabular}{|c|ccccc|ccccc|}
		\hline
		\begin{tabular}[c]{@{}c@{}}Measurement\\ Campaign\end{tabular} & \multicolumn{5}{c|}{RMS delay spread ($ \mu $, ns)} & \multicolumn{5}{c|}{RMS delay spread ($ \sigma $, ns)} \\ \hline
		& \multicolumn{1}{c|}{Meas.} & \multicolumn{1}{c|}{\begin{tabular}[c]{@{}c@{}}Pred. \\ (log. Opt.)\end{tabular}} & \multicolumn{1}{c|}{\begin{tabular}[c]{@{}c@{}}$\Delta$ \\ (log Opt.)\end{tabular}} & \multicolumn{1}{c|}{\begin{tabular}[c]{@{}c@{}}Pred. \\ (lin. Opt.)\end{tabular}} & \begin{tabular}[c]{@{}c@{}}$\Delta$ \\ (lin. Opt.)\end{tabular} & \multicolumn{1}{c|}{Meas.} & \multicolumn{1}{c|}{\begin{tabular}[c]{@{}c@{}}Pred. \\ (log. Opt.)\end{tabular}} & \multicolumn{1}{c|}{\begin{tabular}[c]{@{}c@{}}$\Delta$ \\ (log Opt.)\end{tabular}} & \multicolumn{1}{c|}{\begin{tabular}[c]{@{}c@{}}Pred. \\ (lin. Opt.)\end{tabular}} & \begin{tabular}[c]{@{}c@{}}$\Delta$ \\ (lin. Opt.)\end{tabular} \\ \hline
		28 GHz Outdoor & \multicolumn{1}{c|}{70.4} & \multicolumn{1}{c|}{55.8} & \multicolumn{1}{c|}{14.6} & \multicolumn{1}{c|}{54.7} &  15.7 & \multicolumn{1}{c|}{66.2} & \multicolumn{1}{c|}{66.1} & \multicolumn{1}{c|}{0.1} & \multicolumn{1}{c|}{64.1}& 2.1\\ \hline
		73 GHz Outdoor & \multicolumn{1}{c|}{53.1} & \multicolumn{1}{c|}{53.3} & \multicolumn{1}{c|}{-0.2} & \multicolumn{1}{c|}{52.7} &0.4 & \multicolumn{1}{c|}{56.9} & \multicolumn{1}{c|}{57.8} & \multicolumn{1}{c|}{-0.9} & \multicolumn{1}{c|}{59.2} & -2.3\\ \hline
		140 GHz Outdoor & \multicolumn{1}{c|}{22.7} & \multicolumn{1}{c|}{23.1} & \multicolumn{1}{c|}{-0.4} & \multicolumn{1}{c|}{19.4} &3.3 & \multicolumn{1}{c|}{57.8} & \multicolumn{1}{c|}{51.6} & \multicolumn{1}{c|}{6.2} & \multicolumn{1}{c|}{46.5} &11.3 \\ \hline
	\end{tabular}
	\label{tbl:stats_outdoor_DS}
\end{table*}

\renewcommand{\arraystretch}{1.2}
\begin{table*}[]
	\caption{\centering Difference ($\Delta$) between the RMS angular spread measured (Meas.) using channel measurements and predicted (Pred.) by NYURay in the factory environments at 142 GHz using measurement data reported in \cite{Ju_2022b}.}
	\centering
	\small
	\begin{tabular}{|c|ccccc|ccccc|}
		\hline
		\begin{tabular}[c]{@{}c@{}}Measurement\\ Campaign\end{tabular} & \multicolumn{5}{c|}{RMS angular spread ($ \mu $, ns)} & \multicolumn{5}{c|}{RMS angular spread ($ \sigma $, ns)} \\ \hline
		& \multicolumn{1}{c|}{Meas.} & \multicolumn{1}{c|}{\begin{tabular}[c]{@{}c@{}}Pred. \\ (log. Opt.)\end{tabular}} & \multicolumn{1}{c|}{\begin{tabular}[c]{@{}c@{}}$\Delta$ \\ (log Opt.)\end{tabular}} & \multicolumn{1}{c|}{\begin{tabular}[c]{@{}c@{}}Pred. \\ (lin. Opt.)\end{tabular}} & \begin{tabular}[c]{@{}c@{}}$\Delta$ \\ (lin. Opt.)\end{tabular} & \multicolumn{1}{c|}{Meas.} & \multicolumn{1}{c|}{\begin{tabular}[c]{@{}c@{}}Pred. \\ (log. Opt.)\end{tabular}} & \multicolumn{1}{c|}{\begin{tabular}[c]{@{}c@{}}$\Delta$ \\ (log Opt.)\end{tabular}} & \multicolumn{1}{c|}{\begin{tabular}[c]{@{}c@{}}Pred. \\ (lin. Opt.)\end{tabular}} & \begin{tabular}[c]{@{}c@{}}$\Delta$ \\ (lin. Opt.)\end{tabular} \\ \hline
		Factory A & \multicolumn{1}{c|}{16.3} & \multicolumn{1}{c|}{19.3} & \multicolumn{1}{c|}{-3.0} & \multicolumn{1}{c|}{15.7} & 0.6 & \multicolumn{1}{c|}{20.9} & \multicolumn{1}{c|}{26.6} & \multicolumn{1}{c|}{-5.7} & \multicolumn{1}{c|}{26.4} & -5.5 \\ \hline
		Factory B & \multicolumn{1}{c|}{19.0} & \multicolumn{1}{c|}{16.5} & \multicolumn{1}{c|}{2.5} & \multicolumn{1}{c|}{16.0} &3.0 & \multicolumn{1}{c|}{22.3} & \multicolumn{1}{c|}{22.0} & \multicolumn{1}{c|}{0.3} & \multicolumn{1}{c|}{23.7} &-1.4 \\ \hline
		Factory C & \multicolumn{1}{c|}{25.9} & \multicolumn{1}{c|}{17.8} & \multicolumn{1}{c|}{8.1} & \multicolumn{1}{c|}{18.0} &7.9 & \multicolumn{1}{c|}{29.8} & \multicolumn{1}{c|}{25.7} & \multicolumn{1}{c|}{4.1} & \multicolumn{1}{c|}{30.4} & -0.6\\ \hline
		\multicolumn{1}{|c|}{Factory D} & \multicolumn{1}{c|}{18.5} & \multicolumn{1}{c|}{10.7} & \multicolumn{1}{c|}{7.8} & \multicolumn{1}{c|}{9.9} & \multicolumn{1}{c|}{8.6} & \multicolumn{1}{c|}{27.7} & \multicolumn{1}{c|}{17.2} & \multicolumn{1}{c|}{10.5} & \multicolumn{1}{c|}{17.2} & \multicolumn{1}{c|}{10.5} \\ \hline
	\end{tabular}
	\label{tbl:stats_Factory_AS}
\end{table*}

\renewcommand{\arraystretch}{1.2}
\begin{table*}[]
	\caption{\centering Difference ($\Delta$) between the RMS delay spread measured (Meas.) using channel measurements and predicted (Pred.) by NYURay in the factory environments at 142 GHz using measurement data reported in \cite{Ju_2022b}.}
	\centering
	\small
	\begin{tabular}{|c|ccccc|ccccc|}
		\hline
		\begin{tabular}[c]{@{}c@{}}Measurement\\ Campaign\end{tabular} & \multicolumn{5}{c|}{RMS delay spread ($ \mu $, ns)} & \multicolumn{5}{c|}{RMS delayspread ($ \sigma $, ns)} \\ \hline
		& \multicolumn{1}{c|}{Meas.} & \multicolumn{1}{c|}{\begin{tabular}[c]{@{}c@{}}Pred. \\ (log. Opt.)\end{tabular}} & \multicolumn{1}{c|}{\begin{tabular}[c]{@{}c@{}}$\Delta$ \\ (log Opt.)\end{tabular}} & \multicolumn{1}{c|}{\begin{tabular}[c]{@{}c@{}}Pred. \\ (lin. Opt.)\end{tabular}} & \begin{tabular}[c]{@{}c@{}}$\Delta$ \\ (lin. Opt.)\end{tabular} & \multicolumn{1}{c|}{Meas.} & \multicolumn{1}{c|}{\begin{tabular}[c]{@{}c@{}}Pred. \\ (log. Opt.)\end{tabular}} & \multicolumn{1}{c|}{\begin{tabular}[c]{@{}c@{}}$\Delta$ \\ (log Opt.)\end{tabular}} & \multicolumn{1}{c|}{\begin{tabular}[c]{@{}c@{}}Pred. \\ (lin. Opt.)\end{tabular}} & \begin{tabular}[c]{@{}c@{}}$\Delta$ \\ (lin. Opt.)\end{tabular} \\ \hline
		Factory A & \multicolumn{1}{c|}{14.3} & \multicolumn{1}{c|}{15.3} & \multicolumn{1}{c|}{-1.0} & \multicolumn{1}{c|}{10.9} & 3.4& \multicolumn{1}{c|}{24.2} & \multicolumn{1}{c|}{26.5} & \multicolumn{1}{c|}{-2.3} & \multicolumn{1}{c|}{26.4} & -2.2\\ \hline
		Factory B & \multicolumn{1}{c|}{12.1} & \multicolumn{1}{c|}{13.4} & \multicolumn{1}{c|}{-1.3} & \multicolumn{1}{c|}{13.1} &-1.0 & \multicolumn{1}{c|}{14.5} & \multicolumn{1}{c|}{14.9} & \multicolumn{1}{c|}{-0.4} & \multicolumn{1}{c|}{13.3} &1.2 \\ \hline
		Factory C & \multicolumn{1}{c|}{9.1} & \multicolumn{1}{c|}{5.0} & \multicolumn{1}{c|}{4.1} & \multicolumn{1}{c|}{4.1} & 5.0& \multicolumn{1}{c|}{8.2} & \multicolumn{1}{c|}{6.0} & \multicolumn{1}{c|}{1.8} & \multicolumn{1}{c|}{3.7} & 4.5\\ \hline
		\multicolumn{1}{|c|}{Factory D} & \multicolumn{1}{c|}{22.2} & \multicolumn{1}{c|}{12.9} & \multicolumn{1}{c|}{9.3} & \multicolumn{1}{c|}{11.9} & \multicolumn{1}{c|}{10.3} & \multicolumn{1}{c|}{26.1} & \multicolumn{1}{c|}{20.0} & \multicolumn{1}{c|}{6.1} & \multicolumn{1}{c|}{20.6} & \multicolumn{1}{c|}{5.5} \\ \hline
	\end{tabular}
	\label{tbl:stats_Factory_DS}
\end{table*}

Using the optimization technique in the decibel domain, a good agreement between the measured and simulated multipath powers was observed for the indoor measurement campaigns at 28 and 142 GHz, the outdoor measurement campaigns at 28, 73, and 142 GHz, and for the four factory measurement campaigns (Factory A, Factory B, Factory C, and Factory D - all at 142 GHz), as seen in Fig. \ref{fig:power_error}. 

To determine the probability distribution of the error in the predicted directional multipath power, Akaike’s Information Criterion (AIC) of the Normal, Beta, Gamma, Exponential, Nakagami, Rayleigh, and Rician distribution was compared. The probability distribution within minimal AIC was chosen to model the error in the predicted directional multipath power \cite{Cavanaugh_2019}. The error distribution of the directional multipath component power was a Gaussian random variable in all the measurement campaigns. For the indoor office measurements at 28 and 142 GHz, the mean of the error in power predicted was 0.1 and 0.7 dB, while the standard deviation of the error in power predicted was 4.4 dB and 3.9 dB respectively. For the outdoor environments the mean error was 1.1 dB, 0.0 dB, and 0.0 dB and the standard deviation was equal to 3.2 dB, 2.6 dB and 2.6 dB at 28, 73, and 142 GHz respectively. For the factory environment measurements conducted at 142 GHz, a mean error of 0.5 dB, 0.0 dB, 0.0 dB, and 0.2 dB was measured and the standard deviations of error was equal to 2.3 dB, 3.4 dB, 2.5 dB, and 3.6 dB respectively at Factory A, B, C, and D.

The accuracy of the calibrated ray tracer was characterized by noting the standard deviation of the absolute error in the simulated directional multipath power. For the indoor office measurement campaigns at 28 and 142 GHz, the standard deviation of the absolute error in power predicted was 2.7 dB and 2.8 dB respectively. Since the outdoor environments were less cluttered, NYURay was able to better predict the directional multipath power, with standard deviations of absolute error of 2.0 dB, 1.6 dB and 1.7 dB at 28, 73, and 142 GHz respectively. For the factory environment measurements at 142 GHz, the detailed 360-degree videos of the environment near the TX and RX enabled calibration to greater accuracy than the indoor office measurement campaigns, resulting in standard deviations of absolute error equal to 2.0 dB, 1.9 dB, 1.7 dB, and 2.3 dB respectively at Factory A, B, C, and D.

In Fig. \ref{fig:directional_pdp}, a comparison of a sample PDP created by NYURay with a sample PDP measured during the 28 GHz indoor office measurement campaign is presented. Although NYURay does not predict all of the multipath components that are measured at a delay of 35 ns, there is often good agreement in the other multipath components' powers. The strongest multipath component's measured power, in particular, is within 2 dB of the power that NYURay predicted.

    \captionof{table}{\centering Reflection and penetration loss of common building materials obtained from NYURay calibration at 28, 73, and 142 GHz. The reflection and penetration loss increases with increase in frequency.}
	\begin{supertabular}{|p{2 cm}|p{0.7cm}|p{1.9cm}|p{0.8cm}|p{0.8cm}|}
		\hline
		\textbf{Material}                    &\textbf{$ f_c $}\textbf{ (GHz)} & \textbf{Env.} & \begin{tabular}[c]{@{}c@{}}\textbf{Refl.} \\ \textbf{Loss}\\\textbf{ (dB)}\end{tabular} & \begin{tabular}[l]{@{}l@{}}\hspace{-0.4em}\textbf{Pen.} \\ \textbf{Loss} \\\textbf{(dB)}\end{tabular} \\ \hline
		
		{\begin{tabular}{@{}l@{}}Drywall \cite{Mac15b}\\\cite{Ju19a} \cite{Ju_2022b,Ju_2024a}\end{tabular}}      &    28 & Indoor Office       & 6.1                                                       &4.0   \\ \cline{2-5}
		&    142 & Indoor Office       & 9.9                                                       &9.2   \\ \cline{2-5}
		& 142    &    Factory  A     & 8.7                                                        &13.1    \\ \cline{2-5}
		&142     &    Factory  B     & 12.8                                                        &12.7    \\\cline{2-5}
		&142    &    Factory  C     & 7.9                                                        &-    \\ \cline{2-5}
		&142    &    Factory  D     & 10.1	                                                        &8.0   \\ \hline
		
	{	\begin{tabular}{@{}l@{}}Glass\end{tabular} }     &    28 & Indoor Office       & 3.5                                                       & 3.2   \\ \cline{2-5}
		{	\begin{tabular}{@{}l@{}}\cite{Mac15b,Mac14a,Ju19a}\end{tabular} } 	&73&Outdoor&5.9 &5.0 \\ \cline{2-5}
		{	\begin{tabular}{@{}l@{}}\cite{Xing21a, Shakya_2023,Ju_2022b}\end{tabular} } 	&    142 & Indoor Office       & 24.5                                                       & 7.2   \\ \cline{2-5}
		{	\begin{tabular}{@{}l@{}}\cite{Ju_2024a,MacCartney_2019}\end{tabular} } 	&142&Outdoor&7.4 &3.9 \\ \cline{2-5}
		& 142    &    Factory  A     & 9.9                                                        &10.4    \\ \cline{2-5}						
		&142    &    Factory  D     & 6.9                                                        &-    \\ \hline
		\pagebreak
		Thick Glass \cite{Ju_2022b,Ju_2024a}	& 142    &    Factory  A     & 8.4                                                       &23.0   \\ \hline						
		{	\begin{tabular}{@{}l@{}}Cubicles\\(Fabric)\end{tabular} }      &    28 & Indoor Office       & 3.3                                                       &- \\ \cline{2-5}
		{	\begin{tabular}{@{}l@{}}\cite{Mac15b,Ju19a}\end{tabular} } 	&    142 & Indoor Office       & 8.0                                                    &7.8   \\ \hline
		
		{	\begin{tabular}[c]{@{}l@{}}Wooden\\Cupboard\end{tabular} }      &    28 & Indoor Office       & 3.5                                                       &2.4   \\ \cline{2-5}
	{	\begin{tabular}[c]{@{}l@{}}\cite{Mac15b,Ju19a}\end{tabular} } 	&    142 & Indoor Office       & 0.5                                                       &6.1   \\ \hline				
		{	\begin{tabular}{@{}l@{}}Display Board\end{tabular}  }    &    28 & Indoor Office       & 1.1                                                       &11   \\ \cline{2-5}
	\cite{Mac15b,Ju19a}	&    142 & Indoor Office       & 8.9                                                       &19.1   \\ \hline						
		{\begin{tabular}{@{}l@{}}Whiteboard\end{tabular}}      &    28 & Indoor Office       & 8.3                                                       &11   \\ \cline{2-5}
	\cite{Mac15b,Ju_2022b,Ju_2024a}	&142    &    Factory  D     & -       &8.5    \\ \hline	
		{\begin{tabular}{@{}l@{}}Cardboard Box\\\cite{Mac15b,Ju_2022b,Ju_2024a}\end{tabular} } &    142 & Factory C       & 4.1                                                       &1.7   \\ \hline
	{\begin{tabular}{@{}l@{}}Cork Board    \\\cite{Mac15b,Ju_2022b,Ju_2024a}\end{tabular}}	  
		&    142 & Factory C       &15.3 &-  \\ \hline   
			{\begin{tabular}{@{}l@{}}Wood  \\  \cite{Mac15b,Ju_2022b,Ju_2024a}\end{tabular} }& 142 &Factory D & 4.8 & -\\ \hline
		{\begin{tabular}{@{}l@{}}Cement Wall    \\\cite{Rap13a, Rappaport_2015}\end{tabular}} 	 & 28 &Outdoor & 11.6&-\\ \hline
		{	\begin{tabular}{@{}l@{}}Granite\end{tabular}} & 28 &Outdoor & 6.9&- \\ \cline{2-5}
{	\begin{tabular}{@{}l@{}}\cite{Rap13a, Rappaport_2015,Mac14a}\end{tabular}}		& 73 &Outdoor & 5.6&- \\ \cline{2-5}
{	\begin{tabular}{@{}l@{}}	\cite{Xing21a,Shakya_2023,MacCartney_2019}\end{tabular}}		& 142 &Outdoor &13.1&- \\ \hline
		\pagebreak
		{\begin{tabular}{@{}l@{}}Concrete Pillar  \end{tabular}} & 73 &Outdoor & 12.7&- \\ \cline{2-5}
		{	\begin{tabular}{@{}l@{}} \cite{Mac14a,Xing21a,Shakya_2023} \end{tabular}}	& 142 &Outdoor & 10.3&- \\ \hline
	{	\begin{tabular}{@{}l@{}}Brick Wall\end{tabular}	}   & 73&Outdoor&12.8&-\\ \cline{2-5}
	{	\begin{tabular}{@{}l@{}}\cite{Mac14a,Xing21a,Shakya_2023}  \end{tabular}}	& 142&Outdoor&18.9&-\\ \hline
		{	\begin{tabular}{@{}l@{}}Foliage \end{tabular}}
			 &73&Outdoor&-&6.1\\ \cline{2-5}
	{	\begin{tabular}{@{}l@{}}\cite{Mac14a}\cite{Xing21a}\cite{Shakya_2023}  \end{tabular}}	&142&Outdoor&-&4.6\\ \hline
	\end{supertabular}
	\label{tab:My_values}

\section{Material Properties}

mmWave and sub-THz signals may be reflected, scattered, or may penetrate through materials in the environment. By characterizing reflection, transmission and scattering interactions of a variety of common building materials, signal propagation may be accurately predicted, which will assist in the development of map-generation algorithms. A wide range of materials were present in the diverse measurement environments, such as drywall, glass, granite, wooden cupboards, foliage, and metallic factory equipment.

The channel measurements used to calibrate NYURay provide insight into the reflection and penetration losses of the materials at 28, 73, and 142 GHz as shown in Table \ref{tab:My_values}. The penetration loss of objects was found to increase with increase in frequency. The reflection and penetration loss of drywall and glass obtained from ray tracer calibrations at 28 and 142 GHz agrees well with the channel measurements conducted in \cite{Olsson_2021}. Lower foliage loss was observed at 142 GHz compared to 73 GHz, possibly because the 142 GHz outdoor measurements were conducted in the fall of 2021, when the trees had fewer leaves, while the 73 GHz outdoor measurements were conducted in the summer of 2016. A general trend of increasing reflection loss was observed, however a slight decrease of reflection loss was observed for granite pillars at 73 GHz. 

The reflection loss of materials observed in this article is greater than the reflection loss recommended for common building materials in \cite{ITU-R_P.2040-3}. As per ITU-R P.2040-3, the reflection loss of concrete for normal incidence is 8.1 dB, which is close to the reflection loss of 10.3 dB at 142 GHz and 12.7 dB at 73 GHz observed via calibration in this article. A lower reflection loss of 4.7 dB was observed in \cite{telecom2019analysis}. The reflection loss of brick wall was observed to be 12.7 dB at 73 GHz and 18.9 dB at 142 GHz, higher than the ITU recommendation of 9.7 dB. 

The reflection coefficient of drywall obtained through the calibration procedure varied from 6.1 dB at 28 GHz to 12.8 dB at 142 GHz, in a variety of indoor office and factory environments, with significant variation with location, while a constant reflection loss of 12.1 dB was recommended by ITU. The reflection loss of drywall was shown to vary between 4.8 dB at 60 GHz, 8.2 dB at 73 GHz, and 4.2 dB at 83 GHz in \cite{Khatun_2021}, while a reflection loss of 10.2 dB was observed at 28 GHz in \cite{telecom2019analysis}. Work in \cite{telecom2019analysis} observed a penetration loss of 1.0-2.5 dB for drywall at 28 GHz, while in \cite{Du_2021} a penetration loss of 1.1 dB was observed at 28 GHz while a penetration loss of 3.8 dB was observed at 144 GHz.

A wide variation of measured reflection loss was obtained from the calibration procedure for glass (3.5-24.5 dB), while the ITU recommended reflection loss is 7.3 dB. In \cite{telecom2019analysis}, the reflection losses for thin glass was 10.2 dB and the reflection loss for thick glass was 0.4 dB. A penetration loss of 3 dB and 27.4 dB for two types of glass was observed at 28 GHz in \cite{telecom2019analysis}, which matches well with the penetration loss of 3.2 dB and 23.0 dB obtained from the calibration procedure. In \cite{Du_2021} a lower penetration loss of 0.8 dB and 4.9 dB was observed at 28 and 144 GHz, respectively. Thus, a wide variation of reflection loss and penetration loss is observed in literature, possibly due to subtle differences in materials under test.

\section{Conclusion}

In this study, NYURay, a 3-D mmWave and sub-THz ray tracer, was introduced and calibrated using mmWave and sub-THz channel measurements at 28, 73, and 142 GHz taken in indoor office, outdoor UMi, and factory environments. Accurately calibrated ray tracers can be used to test radio coverage when deploying wireless systems indoors and outdoors, to reduce the chance of coverage gaps, and can generate synthetic data to evaluate sensing algorithms or train machine learning applications, which would otherwise require a large amount of measured data. Traditional calibration approaches assume angle-dependent reflection, requiring slow iterative optimization techniques with no closed form solution. Linear and logarithmic calibration procedures were described in this paper. Both calibration processes performed similarly, however the logarithmic calibration procedure introduced in this paper can quickly and accurately calibrate any ray tracer to real world measurements, providing a closed form solution. Any ray tracer could be easily calibrated to directional data using the calibration process described in this paper. Using the logarithmic calibration procedure, the error in the predicted multipath power was a Gaussian random variable. The calibration performance was characterized by noting the standard deviation of the absolute error of the predicted directional multipath power. In the indoor office environment at 28 and 142 GHz, the standard deviation of absolute error in the multipath power was 2.7 and 2.8 dB, respectively, and 2.0 dB, 1.6 dB, and 1.7 dB in the outdoor environments at 28, 73, and 142 GHz. In the factory environments, due to the availability of a 360 degree video of the environment, the standard deviation in the absolute error of the predicted multipath power was low (2.0 dB, 1.9 dB, 1.7 dB, and 2.3 dB in Factory A, B, C, and D).

 Due to the incompleteness of the environmental map utilized for ray tracing, the RMS angular spread of omnidirectional measurements were underestimated in indoor office setting and the outdoor environment by 10\% while the RMS delay spread was underestimated by 20\%. The precision of mmWave and sub-THz ray tracing can be increased by using a more precise environmental map.
The electrical characteristics of building materials were also reported in this research based on the ray tracer calibration process. The loss due to reflection and penetration increased with frequency. At 28 GHz, materials were the most reflecting when compared to higher frequencies.

\ifCLASSOPTIONcaptionsoff
  \newpage
\fi

\bibliographystyle{IEEEtran}
\bibliography{references,Ray_Tracing}{}

\vskip 0pt plus -1fil

For biographies of authors please refer to: D. Shakya, S. Ju, O. Kanhere, H. Poddar, Y. Xing and T. S. Rappaport, "Radio Propagation Measurements and Statistical Channel Models for Outdoor Urban Microcells in Open Squares and Streets at 142, 73, and 28 GHz," in IEEE Transactions on Antennas and Propagation, vol. 72, no. 4, pp. 3580-3595, April 2024.

\vskip 0pt plus -1fil




\end{document}